\documentclass[11pt]{article}

\usepackage{filecontents}

\usepackage{authblk}

\usepackage[ruled,vlined]{algorithm2e}

\usepackage{clipboard}
\newclipboard{review}

\usepackage{amsthm}
\newtheorem{theorem}{Theorem}
\newtheorem{corollary}{Corollary}
\newtheorem{proposition}{Proposition}

\usepackage{subfigure}
\usepackage{caption}
\usepackage{color}
\usepackage{physics}
\usepackage{hyperref}
\usepackage{graphicx}
\usepackage{amssymb,amsmath}
\usepackage{bm}
\usepackage{color}
\usepackage[authoryear,semicolon]{natbib}
\usepackage{multirow}
\usepackage{booktabs}
\usepackage{array}
\usepackage[normalem]{ulem}
\usepackage{tikz}
\usepackage{siunitx}
\usepackage{url}
\usepackage{standalone}

\newcolumntype{x}[1]{>{\centering\arraybackslash\hspace{0pt}}p{#1}}

\newcommand{\E}{\text{E}}
\newcommand{\Var}{\text{var}}

\newcommand{\cov}{\text{cov}}
\newcommand{\bdiag}{\text{bdiag}}

\newcommand{\sign}{\text{\normalfont sign}}

\newcommand{\K}{\text{K}}
\newcommand{\betavec}{\boldsymbol{\beta}}
\newcommand{\epsilonvec}{\boldsymbol{\epsilon}}
\newcommand{\thetavec}{\boldsymbol{\theta}}

\newcommand{\Sigmavec}{\boldsymbol{\Sigma}}
\newcommand{\sigmavec}{\boldsymbol{\sigma}}

\newcommand{\Pivec}{\boldsymbol{\Pi}}
\newcommand{\zerovec}{\mathbf{0}}

\newcommand{\Cvec}{\mathbf{C}}
\newcommand{\Cmat}{\mathbf{C}}
\newcommand{\mub}{\boldsymbol{\mu}}

\newcommand{\bvec}{\mathbf{b}}
\newcommand{\dvec}{\mathbf{d}}
\newcommand{\fvec}{\mathbf{f}}
\newcommand{\gvec}{\mathbf{g}}
\newcommand{\hvec}{\mathbf{h}}

\newcommand{\Gau}{\text{Gau}}

\newcommand{\Lvec}{\mathbf{L}}

\newcommand{\Avec}{\mathbf{A}}
\newcommand{\avec}{\mathbf{a}}
\newcommand{\ovec}{\mathbf{o}}
\newcommand{\svec}{\mathbf{s}}
\newcommand{\uvec}{\mathbf{u}}

\newcommand{\xvec}{\mathbf{x}}
\newcommand{\Vvec}{\mathbf{V}}

\newcommand{\Xvec}{\mathbf{X}}
\newcommand{\Yvec}{\mathbf{Y}}
\newcommand{\Zvec}{\mathbf{Z}}
\newcommand{\Yt}{\tilde{Y}}
\newcommand{\Yvect}{\tilde{\mathbf{Y}}}

\let\originalleft\left
\let\originalright\right
\renewcommand{\left}{\mathopen{}\mathclose\bgroup\originalleft}
\renewcommand{\right}{\aftergroup\egroup\originalright}
\renewcommand{\arraystretch}{1.2}

\usepackage{xr}
\usepackage{geometry}
\begin{document}

\def\spacingset#1{\renewcommand{\baselinestretch}%
{#1}\small\normalsize} \spacingset{1.5}


  \title{\bf 
  Modeling Nonstationary and Asymmetric Multivariate Spatial Covariances via Deformations}
  \author[1, *]{Quan Vu}
  \author[1]{Andrew Zammit-Mangion}
  \author[1]{Noel Cressie}
  \affil[1]{School of Mathematics and Applied Statistics, University of Wollongong, Australia}
  \affil[*]{Corresponding author: quanv@uow.edu.au}
  \date{}
  \maketitle

\smallskip

\begin{abstract}
Multivariate spatial-statistical models are often used when modeling environmental and socio-demographic processes. The most commonly used models for multivariate spatial covariances assume both stationarity and symmetry for the cross-covariances, but these assumptions are rarely tenable in practice. In this article we introduce a new and highly flexible class of nonstationary and asymmetric multivariate spatial covariance models that are constructed by modeling the simpler and more familiar stationary and symmetric multivariate covariances on a warped domain. Inspired by recent developments in the univariate case, we propose modeling the warping function as a composition of a number of simple injective warping functions in a deep-learning framework. Importantly, covariance-model validity is guaranteed by construction. We establish the types of warpings that allow for cross-covariance symmetry and asymmetry, and we use likelihood-based methods for inference that are computationally efficient. The utility of this new class of models is shown through two data illustrations: a simulation study on nonstationary data and an application on ocean temperatures at two different depths.
\end{abstract}

\noindent%

\spacingset{1.5} 

\def\thefigure{\arabic{figure}}
\def\thetable{\arabic{table}}

\renewcommand{\theequation}{\thesection.\arabic{equation}}

\setcounter{section}{0} 
\setcounter{equation}{0} 

\section{Introduction ~}\label{sec:Intro}

Multivariate spatial-statistical models are used to jointly model two or more variables that are spatially indexed. They find widespread use in several application domains, such as the environmental sciences and the social sciences, where spatial processes interact. The utility of multivariate models lies in the concept of `borrowing strength,' where in this setting information on one process (obtained, for example, through observation) imparts information on the other processes that are being jointly modeled but that may or may not be directly observed. Multivariate spatial models need to adequately model both the marginal behavior of the spatial processes as well as the joint dependence between the processes. Often, the central object of interest when constructing a multivariate spatial model is the cross-covariance matrix function, which encodes the marginal covariances and cross-covariances of the spatial processes and its use improves inference over one-at-a-time univariate analyses of each process.

Typically, the two simple assumptions of stationarity and symmetry are made when modeling the marginal behavior of, and the joint dependence between, multiple processes. For example, the popular linear model of coregionalization (LMC) \citep[e.g.,][]{goulard1992linear, wackernagel2003multivariate} assumes both of these properties, as does a recent multivariate model constructed via spectral representations \citep{qadir2019semiparametric}, and the multivariate Mat{\'e}rn model \citep{gneiting2010matern, apanasovich2012valid}. \Copy{rev2p3}{While the multivariate Mat{\'e}rn model, where the elements of the cross-covariance matrix function are all Mat{\'e}rn covariance functions, has proven to be considerably more flexible than the traditional LMC, covariance nonstationarity and asymmetry are present in many scientific applications and should be modeled. For example, the rates of ice loss in Antartica are clearly nonstationary, since more ice loss occurs in regions of high ice-stream velocity, which are at the boundary of the continent \citep{zammit2015multivariate}; and in the inversion of a trace-gas, the cross-covariance between the flux field and the mole-fraction field is asymmetric because of atmospheric transport \citep{zammit2015spatio}. Asymmetry of cross-covariances is clearly present in the ocean-temperatures data in Section \ref{sec:applications_ocean}, due to diffusive and advective oceanographic processes.}

Nonstationarity in a univariate setting has been addressed through the use of spatial deformations \citep[e.g.,][]{sampson1992nonparametric, damian2001bayesian, schmidt2003bayesian2, fouedjio2015estimation}; basis functions \citep[e.g.,][]{cressie2008fixed}; stochastic partial differential equations (SPDEs) \citep[e.g.,][]{lindgren2011explicit, fuglstad2015exploring}; and process convolution with a spatially varying kernel, which leads to spatially varying model parameters \citep[e.g.,][]{higdon1999non, paciorek2006spatial}. Addressing nonstationarity in a multivariate setting is more problematic, as one needs to ensure validity, namely nonnegative-definiteness, of all possible covariance matrices that are constructed through a proposed cross-covariance matrix function. Valid spatial multivariate modeling approaches that account for nonstationarity include those of \cite{gelfand2004nonstationary} and \cite{kleiber2012nonstationary}, who extended the LMC and multivariate Mat{\'e}rn model, respectively, to contain spatially varying parameters. Other approaches consider basis functions \citep{Nguyen_2017} or systems of SPDEs \citep{Hu_2016, hildeman2019joint}. 

Some multivariate models also model asymmetry. \Copy{rev1p9}{For example, \cite{li2011approach} built on the bivariate example given by \cite{ver1993multivariable} and extended the general stationary multivariate model to the asymmetric case}. On the other hand, \cite{apanasovich2010cross} used latent dimensions to model asymmetric cross-covariances, and \cite{cressie2016multivariate} used a non-centred kernel to introduce asymmetry in the joint dependence structure; see also \citet{ver1998constructing} and \citet{majumdar2007multivariate}.

An attractive way to introduce nonstationarity and asymmetry in multivariate spatial-statistical models is through a generalization of the univariate deformation approach of \citet{sampson1992nonparametric}. It is well known that nonstationarity can be modeled by deforming space; specifically, a stationary process on a warped domain can induce a highly nonstationary process on the original (geographic) domain. In the multivariate case, one may apply a common deformation to all of the processes or separate deformations to each process. As we demonstrate in this article, using a common deformation function enforces symmetry and constrains the nonstationary behavior (i.e., the local anisotropies and scales) to be common for each process. However, process-specific deformation functions allow for distinct nonstationary behavior and practically guarantee asymmetry. Multivariate models that are built using spatial deformations bring with them several advantages over some of the other models mentioned above. In particular, they are able to capture complex nonstationary and asymmetric behavior; their cross-covariance functions are valid by construction; and, if deep compositional warping functions are used, they are computationally efficient to fit and predict with.

This article is organized as follows. In Section 2, we first give some background on multivariate spatial models and univariate deep compositional spatial models (DCSMs). In Section 3, we introduce a multivariate generalization of the univariate DCSM and show how asymmetry can be induced in a simple manner through what we call `aligning functions.' In Section 4, we show results from two data illustrations: First, we show the utility of multivariate DCSMs for modeling symmetric nonstationary simulated data; then we show its utility in an application on modeling Atlantic Ocean temperatures at two different depths. In these illustrations of our methodology, we show, through cross-validation and visualization, that spatial predictions from the multivariate DCSMs are generally superior to those from conventional multivariate spatial models. In Section 5, we summarize our conclusions. Additional material is provided in the Supplementary Material.

\section{Background ~}\label{sec:Model}

The multivariate DCSM that we construct in Section~\ref{sec:Model_multi} requires the specification of a conventional symmetric, stationary, possibly isotropic, multivariate covariance model, and a deep warping function. In this section we briefly review these two building blocks.

\subsection{Multivariate Spatial Covariance Models ~}

Consider a $p$-variate spatial process $\Yvec(\svec) \equiv ( Y_1(\svec),\dots, Y_p(\svec))'$, $\svec \in G$, where we refer to $G \subset \mathbb{R}^d$ as the \emph{geographic} domain in $d$-dimensional Euclidean space. We assume that $\Var(Y_i(\svec)) < \infty$, for all $\svec \in G$ and all $i = 1,\dots,p$. Therefore this multivariate process has finite expectation $\mub(\cdot) \equiv (\mu_1(\cdot), \dots, \mu_p(\cdot))'$, and a valid cross-covariance matrix function $\Cmat_G(\cdot \, , \cdot) \equiv (C_{ij,G}(\cdot \, , \cdot): i,j = 1,\dots,p)$, where $C_{ij,G}(\svec, \uvec) = \cov(Y_i(\svec), Y_j(\uvec)); \svec, \uvec \in G$. For $i,j = 1,\dots,p$, the covariance function $C_{ii,G}(\cdot \,, \cdot)$ is the covariance function of the process $Y_i(\cdot)$ and, for $i \neq j$, $C_{ij,G}(\cdot \, , \cdot)$ is the cross-covariance function of $(Y_i(\cdot), Y_j(\cdot))'$.

In some cases, the cross-covariance matrix function only depends on $\hvec \equiv \svec - \uvec$. That is, for $i,j = 1,\dots, p$,
\( C_{ij,G}(\svec, \uvec) \equiv C^{o}_{ij,G}(\hvec); \svec,\uvec \in G, \)
where now each $C^{o}_{ij,G}(\cdot)$ is a function of displacement. In this case, we say that $\Cmat_{G}^{o}(\cdot) = (C^{o}_{ij,G}(\cdot) : i,j = 1,\dots,p) $ is a stationary cross-covariance matrix function. A cross-covariance matrix function is said to be symmetric if, for $i, j = 1, \dots, p,$ $C_{ij,G}(\svec, \uvec) = C_{ji,G}(\svec, \uvec),$ for $\svec,\uvec \in G$. In the stationary case, symmetry is given by $C_{ij,G}^{o}(\hvec) = C_{ji,G}^{o}(\hvec),$ for $\hvec = \svec - \uvec$ and $\svec, \uvec \in G$.

Stationarity and symmetry are strong assumptions in practice, but they remain popular since they facilitate the construction of valid cross-covariance matrix functions with a relatively small set of parameters. Among the most popular stationary, symmetric, multivariate covariance models is the multivariate Mat{\'e}rn model \citep{gneiting2010matern}, where the marginal covariance functions and cross-covariance functions are all Mat{\'e}rn covariance functions. In this model, every process may have a different degree of smoothness, thus circumventing a key limitation of the LMC where for $i= 1,\dots,p$, the smoothness of each $Y_i(\cdot)$ is generally the same by construction.

The isotropic Mat{\'e}rn \textit{correlation} function is given by
\( \mathcal{M}(\hvec | \nu, a) = \frac{2^{1-\nu}}{\Gamma(\nu)} (a \norm{\hvec})^{\nu} \K_{\nu}(a \norm{\hvec}), \)
where $\nu$ is the smoothness parameter, $\K_{\nu}(\cdot)$ is the modified Bessel function of the second kind of order $\nu$, $a$ is the scale parameter, and $\Gamma(\cdot)$ is the gamma function. \Copy{rev3iip1}{A multivariate spatial-statistical process $\Yvec(\cdot)$ has a multivariate Mat{\'e}rn cross-covariance matrix function if, for $i,j = 1,\dots,p$, and $\svec,\uvec \in G$, 
	\begin{align}\label{eq:matern_model}
		\begin{split}
			C_{ii,G}(\svec, \uvec) &\equiv \cov(Y_i(\svec), Y_i(\uvec)) = \sigma_{i}^2 \mathcal{M}(\hvec | \nu_{ii}, a_{ii}), \\
			C_{ij,G}(\svec, \uvec) &\equiv \cov(Y_i(\svec), Y_j(\uvec)) = \rho_{ij} \sigma_{i} \sigma_{j} \mathcal{M}(\hvec | \nu_{ij}, a_{ij}), \quad \text{for } i \neq j,
		\end{split}
	\end{align}
	where $\{a_{ij} \}$ are scale parameters, $\{\nu_{ij} \}$ are smoothness parameters, $\{\sigma_i^2 \}$ are variance parameters, and $\{\rho_{ij}\}$ are cross-correlation parameters. From \eqref{eq:matern_model}, we can see that multivariate Mat{\'e}rn cross-covariance functions are stationary, symmetric, and isotropic cross-covariance functions.} In order to ensure validity, some constraints must be placed on the parameters of the multivariate Mat{\'e}rn covariance models shown in \eqref{eq:matern_model}. The parsimonious Mat{\'e}rn covariance models have even stricter constraints than the more general multivariate Mat{\'e}rn models \citep[see also][]{apanasovich2012valid}, but they have been shown to be flexible enough to model several environmental processes of interest \citep[see][for more details]{gneiting2010matern}.
We shall use the multivariate parsimonious Mat{\'e}rn covariance model in Section~\ref{sec:illustrations} to construct multivariate DCSMs for two bivariate spatial data sets.

\subsection{Deep Compositional Spatial Models ~}

The univariate deep compositional spatial modeling approach of \cite{azm2019deep} uses injective warpings to construct nonstationary covariance models from simple covariance models. The idea to use deformations (or warpings) to modify the properties of a process stems from the work of  \cite{sampson1992nonparametric}; see also \cite{Meiring_1997}, \cite{Sampson_2001}, \cite{schmidt2003bayesian2}, \cite{Calandra_2016}, and references therein. In this article, we extend the univariate deep compositional approach to the important multivariate case.

Consider for the moment a univariate process $Y(\cdot)$ with $\Var(Y(\svec)) < \infty$, $\svec \in G$, and with nonstationary covariance function $C_{G}(\cdot \, , \cdot)$. After warping the space $G$, suppose that $C_G(\cdot \, , \cdot)$ can be expressed as a simpler, stationary, covariance function, $C_{D}(\cdot \, , \cdot)$, on a deformed space $D$, through a warping function $\fvec: G \to D$. Specifically, $C_G(\svec,\uvec) \equiv C_D(\fvec(\svec), \fvec(\uvec))$, for $\svec, \uvec \in G$, where $C_{D}(\cdot \, , \cdot)$ is a familiar (stationary) covariance function. In DCSMs, the warping function $\fvec$ is constrained to be smooth and injective in order to preclude the possibility of space-folding; see also \cite{perrin1999modelling}. In particular, it is expressed as the composition,
\(
\fvec(\cdot) \equiv \fvec_{[L]}\,\circ\,\fvec_{[L-1]}\,\circ\,\cdots\,\circ\,\fvec_{[1]}(\cdot),   
\)
where $\fvec_{[1]}(\cdot),\dots, \fvec_{[L]}(\cdot)$ are simple elemental injective functions, and $L$ is the number of warpings (or layers). This compositional construction is very flexible in that it can model highly nonstationary spatial processes, yet it is simple enough to facilitate parameter estimation from relatively sparse data. \cite{azm2019deep} call the functions $\fvec_{[1]}(\cdot),\dots, \fvec_{[L]}(\cdot)$ warping units, and propose three types: axial warping units, radial basis function units, and M{\"o}bius transformation units. In this article, we also make use of these three types of warping units; see Table \ref{tbl:warping_units} in the Supplementary Material for more details on these units. \Copy{rev3iip4}{For example, in Section \ref{sec:simstudies_sym}, the warping function $\fvec(\cdot)$ is a composition of $L = 4$ warping units, where $\fvec_{[1]}(\cdot)$ and $\fvec_{[2]}(\cdot)$ are two axial warping units (one for each spatial dimension), $\fvec_{[3]}(\cdot)$ is a radial basis function unit, and $\fvec_{[4]}(\cdot)$ is a M{\"o}bius transformation unit.}

\cite{azm2019deep} modeled a low-rank univariate process that was approximately stationary on the warped domain. Here, we are in a multivariate setting, and we construct valid flexible models for covariances and cross-covariances on a geographic domain $G$ by considering a stationary and symmetric cross-covariance matrix function on the warped domain $D$. We use the warping functions to model nonstationary and asymmetric behavior of the multivariate spatial process.

\section{Multivariate Deep Compositional Spatial Models ~}\label{sec:Model_multi}

\subsection{Model ~}\label{sec:Model_multi_main}

We now construct multivariate DCSMs by extending the univariate construction of \cite{azm2019deep}. In the univariate case, one warping function is all that is required; however, in the $p$-variate case, we could use $p$ different warping functions, one for each process.

We start off with the special case where a single warping function is used for all of the $p$ processes. In this case, for $i,j = 1,\dots,p$, we have:
\begin{equation}\label{eq:cov_model}
	C_{ij, G}(\svec, \uvec) = C_{ij, D}(\fvec(\svec), \fvec(\uvec)) = C^{o}_{ij, D}({\fvec(\svec) - \fvec(\uvec)}); \quad \svec,\uvec \in G,
\end{equation}
where $\Cvec^{o}_{D}(\cdot) \equiv (C^{o}_{ij, D}(\cdot): i,j = 1,\dots,p)$ is a stationary, nonnegative-definite cross-covariance matrix function.

\begin{proposition}\label{propo:symm}
	If $\Cvec^{o}_{D}(\cdot)$ is symmetric, then the cross-covariance matrix function $\Cmat_G(\cdot \, , \cdot)$ defined in \eqref{eq:cov_model}, is symmetric.
\end{proposition}

\noindent \textit{Proof}: See Section \ref{sec:propo:symm_proof} of the Supplementary Material.

Consider now the case where $p$ warpings, one for each process, are used for constructing the cross-covariance matrix function of the $p$-variate process. In this case, for $i,j = 1,\dots,p$,
\begin{equation}\label{eq:cov_model_asym_general}
	C_{ij, G}(\svec, \uvec) = C_{ij, D}(\fvec_i(\svec), \fvec_j(\uvec)) = C^{o}_{ij, D}({\fvec_i(\svec) - \fvec_j(\uvec)}); \quad \svec,\uvec \in G,
\end{equation}
where $\{\fvec_i(\cdot): i = 1,\dots, p\}$ are process-specific warping functions and, as in Proposition \ref{propo:symm}, the valid, stationary cross-covariance matrix function $\Cvec^{o}_{D}(\cdot)$ is symmetric.

\begin{proposition}\label{propo:asymmetry}
	If $\Cvec^{o}_{D}(\cdot)$ is symmetric, then the cross-covariance matrix function $\Cmat_G(\cdot, \cdot)$ defined in \eqref{eq:cov_model_asym_general} is not necessarily symmetric.
\end{proposition}

\noindent \textit{Proof}: See Section \ref{sec:propo:asymmetry_proof} of the Supplementary Material.

The validity of the cross-covariance matrix function constructed via warping functions is established through the following proposition.

\begin{proposition}\label{propo:valid}
	Assume that $\Cmat^o_D(\cdot)$ is a valid, stationary cross-covariance matrix function. Consider the spatial locations $\{ \svec_{i1},\dots,\svec_{i n_i} \}$, where $n_i > 0$, $i = 1,\dots,p$. Let $\Sigmavec_G = (\Sigmavec_{ij, G}: i,j = 1,\dots,p)$, where $\Sigmavec_{ij, G} = (C_{ij, G}(\svec_{ik}, \svec_{jl}): k=1,\dots,n_i, ~ l = 1,\dots,n_j )$, $N = \sum_{i=1}^{p} n_i$, and where $C_{ij,G}(\cdot, \cdot)$ is given by \eqref{eq:cov_model_asym_general}. Then, $\Sigmavec_G$ is nonnegative-definite.
\end{proposition}

\noindent \textit{Proof}: See Section \ref{sec:propo:valid_proof} of the Supplementary Material.

Summarizing the results of Proposition 1--3, we see that if $\Cmat^o_D(\cdot)$ is a valid, stationary cross-covariance matrix function, then the cross-covariance matrix function $\Cmat_{G}(\cdot \, , \cdot)$ constructed through \eqref{eq:cov_model_asym_general} is valid (i.e., nonnegative-definite). Further, if $\fvec_i(\cdot) \ne \fvec_j(\cdot),$ for any $i,j = 1,\dots, p$, then the cross-covariance matrix function is not necessarily symmetric.

\Copy{rev1p3}{Using $p$ general warpings as in \eqref{eq:cov_model_asym_general} will yield a highly flexible, parameterized model, but one that may be prone to over-fitting.} In practice, any asymmetry present is likely to be simple and dominated by global shifts and rotations. Hence, to model asymmetry, we propose expressing each $\fvec_i(\cdot)$ as a composition of a shared warping function $\fvec(\cdot)$, and a process-specific `aligning function' $\gvec_i(\cdot)$, for $i = 1,\dots,p$. That is, for $i,j = 1,\dots,p$, we let
\begin{align}\label{eq:cov_model_asym}
	C_{ij, G}(\svec, \uvec) &= C_{ij, D}(\fvec \circ\gvec_i(\svec), \fvec \circ \gvec_j(\uvec)) \nonumber \\
	&= C^{o}_{ij, D}(\fvec \circ \gvec_i(\svec) - \fvec \circ\gvec_j(\uvec)); \quad \svec,\uvec \in G,
\end{align}
where $\gvec_i(\cdot), i = 1,\dots,p,$ are simple transformations that are commonly used to align spatial fields and which can include translations and rotations \citep[e.g.,][]{wiens2019surface}. It can be seen from Propositions 2 and 3 that the effect of the aligning functions $\{ \gvec_i \}$ is to introduce asymmetry while preserving nonnegative-definiteness. Note that a common frame of reference for the aligning functions needs to be chosen when aligning fields in this way. Therefore, without loss of generality, we fix $\gvec_1(\cdot)$ to be the identity map, in which case $\fvec_1(\cdot) = \fvec \circ \gvec_1(\cdot)$, is simply the shared warping function $\fvec(\cdot)$. \Copy{rev1p10}{The cross-covariance model in \eqref{eq:cov_model_asym} is a generalization of the asymmetric cross-covariance model of \cite{li2011approach}, where the shared warping function $\fvec(\cdot)$ is the identity map and the aligning functions are translations, that is, where $\gvec_i(\svec) = \svec + \dvec_i,$ for some $\dvec_i \in \mathbb{R}^2$, $i = 2,\dots,p$.}

Under the cross-covariance-matrix model \eqref{eq:cov_model_asym}, nonstationarity can be introduced through both the shared warping function $\fvec(\cdot)$ and the aligning functions $\{ \gvec_i(\cdot): i = 1,\dots,p \} $. \Copy{rev1p8}{Perhaps not surprisingly, the aligning functions can induce nonstationarity in the cross-covariance functions, even when $\fvec(\cdot)$ is the identity map, as we demonstrate in the following proposition.}

\begin{proposition}\label{propo:marginal_stat_model}
	Consider the p-variate cross-covariance matrix model \eqref{eq:cov_model_asym} where $\fvec(\cdot)$ is the identity map; one of the aligning functions $\gvec_k(\cdot)$, for some $k \in \{ 2,\dots, p \}$, is an affine transformation, and $\{ \gvec_i(\cdot), i \neq k \}$ are identity maps. Then, $C_{ik,G}(\cdot, \cdot), i \neq k,$ is not necessarily stationary.
\end{proposition}  

\noindent \textit{Proof}: See Section \ref{sec:propo:marginal_stat_model_proof} of the Supplementary Material.

Proposition \ref{propo:marginal_stat_model} represents one simple way to introduce nonstationarity. More generally, when one has $p$ warping functions $\{\fvec_i(\cdot): i = 1,\dots, p\}$, nonstationarity of $\Cvec_{ij,G}(\cdot, \cdot)$ is obtained by choosing $\fvec_i(\cdot)$ and $\fvec_j(\cdot)$ such that $\fvec_i(\svec) - \fvec_j(\uvec)$ is not a function of $\svec - \uvec$ for $\svec, \uvec \in G$.

\subsection{Parameter Estimation ~}\label{sec:param_est}

Assume now that we have observations $\{Z_{ik}: k = 1,\dots, n_i;\ i = 1,\dots,p\}$ of a $p$-variate Gaussian process $\tilde \Yvec(\cdot)$, where
\begin{equation}\label{eq:z_model}
	Z_{ik} = \Yt_i(\svec_{ik}) + \epsilon_{ik}; \quad k = 1,\dots,n_i, \quad i = 1,\dots,p.
\end{equation}
In \eqref{eq:z_model}, $\{\epsilon_{ik}: k = 1,\dots,n_i; i = 1,\dots,p \}$ are independent Gaussian measurement errors that satisfy, $\epsilon_{ik} \sim \Gau(0, \tau_i^2)$ for $k = 1, \dots, n_i$ and $i = 1,\dots,p$; and $\tau_1^2,\dots,\tau_p^2$ are the measurement-error variances that are assumed to be process-specific and hence all potentially different. We model the $p$-variate Gaussian process $\tilde \Yvec(\cdot) = (\tilde Y_1(\cdot),\dots,\tilde Y_p(\cdot))'$ to have first moment that is linear in covariates $\xvec(\cdot) \equiv (x_1(\cdot), x_2(\cdot),\dots, x_q(\cdot))'$. That is,
\begin{equation}\label{eq:y_model}
	\Yt_i(\cdot) = \xvec(\cdot)' \betavec_i + Y_i(\cdot),  \quad i = 1,\dots,p,
\end{equation}
where $\betavec_1,\dots,\betavec_p \in \mathbb{R}^q$, are vectors of unknown coefficients that need to be estimated, and now $\Yvec(\cdot) \equiv (Y_1(\cdot),\dots,Y_p(\cdot))'$ is a zero-mean second-order nonstationary multivariate Gaussian process on the geographic domain $G$. 

Let $\Yvect_i \equiv (\Yt_i(\svec_{i1}),\dots, \Yt_i(\svec_{i n_{i}}))'$, $\Yvect \equiv (\Yvect_1',\dots, \Yvect_p')'$, $\Yvec_i \equiv (Y_i(\svec_{i1}),\dots, Y_i(\svec_{i n_{i}}))'$, $\Yvec \equiv (\Yvec_1',\dots, \Yvec_p')'$, $\Xvec_i \equiv (\xvec(\svec_{i1}),\dots, \xvec(\svec_{i n_{i}}))'$, $\Xvec = \bdiag(\Xvec_1,\dots, \Xvec_p)$, where $\bdiag(\cdot)$ returns a block diagonal matrix from its arguments, and $\betavec = (\betavec_1', \dots, \betavec_p')'$. Then \eqref{eq:y_model} can be written compactly as
\begin{equation}\label{eq:yt_model}
	\Yvect = \Xvec \betavec + \Yvec.
\end{equation}
The covariance matrix of $\Yvec$, $\Sigmavec_{G} \equiv \cov(\Yvec)$, is given by $(\Sigmavec_{ij, G} : i,j = 1,\dots,p)$, where $\Sigmavec_{ij, G} \equiv (C_{ij, G}(\svec_{ik}, \svec_{jl}): k = 1,\dots,n_i; l = 1,\dots,n_j)$. Furthermore, $\cov(\tilde \Yvec) = \cov(\Yvec)$.

Let $\Zvec_i \equiv (Z_{i1},\dots, Z_{i n_{i}})'$, $\Zvec \equiv (\Zvec_1',\dots, \Zvec_p')'$, $\epsilonvec_i \equiv (\epsilon_{i1},\dots \epsilon_{i n_{i}})'$, and $\epsilonvec \equiv (\epsilonvec_1',\dots, \epsilonvec_p')'$. Then from \eqref{eq:z_model} and \eqref{eq:yt_model}, we have,
$ \Zvec = \Xvec \betavec + \Yvec + \epsilonvec,$
where the covariance matrix of $\epsilonvec$, $\Vvec \equiv \cov(\epsilonvec)$, is diagonal. The model for the observations $\Zvec$ is therefore
\begin{equation}\label{eq:gau_model}
	\Zvec \sim \Gau(\Xvec \betavec, \Sigmavec_{Z}),
\end{equation}
where $\Sigmavec_{Z} = \Sigmavec_{G} + \Vvec$.

Likelihood-based inference can be used to estimate the parameters (including warping parameters) in \eqref{eq:gau_model} \citep{azm2019deep}. Here, we use restricted maximum likelihood (REML) to estimate the parameters in $\Sigmavec_{Z}$, since it is known to provide less-biased estimators of variance-component parameters \citep{cressie1996asymptotics}. \Copy{rev3iip3}{Let $\thetavec$ be the vector containing all parameters appearing in the covariance matrix $\Sigmavec_{Z}$, which includes the unknown parameters appearing in the warping function $\fvec(\cdot)$ (and the aligning functions $\{\gvec_i(\cdot)\}$, if present), the parameters in the cross-covariance matrix function of the process on $D$ (i.e., the scale parameter $a$, the smoothness parameters $\{\nu_{ij} \}$, the variance parameters $\{\sigma_i^2 \}$, and the cross-correlation parameters $\{\rho_{ij}\}$, for the parsimonious Mat{\'e}rn covariance function), and the measurement-error variances.} The restricted maximum likelihood estimate $ \hat{\thetavec} $ of $\thetavec$ is found by maximizing the log restricted likelihood $\mathcal{L}(\thetavec; \Zvec)$ with respect to $\thetavec$ for some given $\Zvec$. After obtaining $ \hat{\thetavec} $, the associated estimate $ \hat{\betavec} $ of $\betavec$ is found through generalized least squares. Once we obtain the REML parameter estimates, these are ``plugged in'' and allow spatial predictions of the hidden processes at an unobserved location $\svec^{*}$. For more details on fitting and prediction, see Section \ref{sec:predictions} of the Supplementary Material.

\Copy{complexity}{The computational time complexity of evaluating the restricted likelihood is the sum of that for evaluating the deformation function and that for factorizing the full joint covariance matrix on the deformed space. The computational complexity of evaluating the aligning functions $\{ \gvec_i(\cdot) \}$, when these are affine transformations, is $O(\sum_{i=2}^{p} n_i)$. The computational complexity of evaluating the shared warping-function layer $\fvec_{[l]}(\cdot)$ is $O(N r_l)$, where $N = \sum_{i=1}^{p} n_i$ is the total number of observations for all processes, and $r_l$ is the number of basis functions in $\fvec_{[l]}(\cdot)$. Hence, the total complexity for evaluating the deformation function is $O(\sum_{i=2}^{p} n_i + N \sum_{l=1}^{L} r_l)$. The complexity of factorizing the covariance matrix on the deformed space is $O(N^3)$. Usually we choose $r_l \ll N,$ for $l = 1, \dots, L$, so that the factorization of the covariance matrix on the deformed space dominates. The actual runtime also changes with the number of iterations used for optimizing the parameter estimates, which needs to be larger when the model is more complex.}
\Copy{rev3p1}{Parameter estimation was done using gradient-based optimization via the \texttt{R} package \texttt{tensorflow} \citep{TensorflowR}, which computes gradients using automatic differentiation, and which can be run on a graphics processing unit (GPU).} 

\subsection{Fixing the Frame of Reference ~}\label{sec:frame_of_ref}

While setting $\gvec_1(\cdot)$ to be the identity map establishes a common frame of reference for the aligning functions, the shared warping function $\fvec(\cdot)$ and any scale parameters appearing in the cross-covariance matrix function are themselves non-identifiable unless this common frame of reference is fixed. Non-identifiability occurs when there exists at least two distinct parameters, $\thetavec_1$, $\thetavec_2$ say, for which $\mathcal{L}(\thetavec_1; \cdot) = \mathcal{L}(\thetavec_2; \cdot)$ \citep{kadane1974role}. \Copy{rev3iip8}{Note that this notion of non-identifiability of parameters is different from the identifiability problem of consistent estimation under infill asymptotics described by  \cite{zhang2004inconsistent}.}
In our case, if we use a stationary, symmetric, cross-covariance matrix function on the warped domain that is also isotropic, the likelihood is invariant to translation, rotation, and reflection of $\fvec(\cdot)$. Since we also allow $\fvec(\cdot)$ to stretch and contract the geographic domain, any scale parameter $a_{ij}$ associated with the cross-covariance function $C_{ij,D}^{o}(\cdot)$ is also non-identifiable \citep[see][for more details]{anderes2008estimating, anderes2009consistent}. While this invariance and lack of identifiability does not pose a problem for prediction, it does mean that we cannot make inference on certain properties of the warping function, such as stretches/contractions and rotations, without further assumptions. As we discuss next, it helps to use a function $\bvec_0(\cdot)$, which we call a \textit{homogenizing function}, to place the estimates of $\fvec(\cdot)$ in a fixed frame of reference and to obtain transformations of the scale parameters that are identifiable.

We illustrate our methodology on the two-dimensional Euclidean space with $d = 2$. Specifically, we establish a fixed frame of reference (which can be easily generalized for $d > 2$), when we assume isotropy on $D$, as follows. Consider three locations, say $\svec_k$, $\svec_l$, and $\svec_m$ in $G \subset \mathbb{R}^2$ such that $\fvec(\svec_k)$, $\fvec(\svec_l)$, and $\fvec(\svec_m)$ are not colinear. Then we use the homogenizing function to shift, scale, rotate, and reflect the warped domain $D$ such that $\bvec_0\circ\fvec(\svec_k) = (0,0)'$,  $\bvec_0\circ\fvec(\svec_l) = (1,0)'$, and \Copy{rev3ip4b}{$b_{0,2}\circ\fvec(\svec_m) > 0$, where $b_{0,2}(\cdot)$ refers to the second element of $\bvec_0(\cdot)$.} A homogenizing function that accomplishes these transformations is given by\begin{equation}\label{eq:b_func}
	\bvec_0(\cdot) \equiv \bvec_3 \circ \bvec_2 \circ \bvec_1(\cdot),
\end{equation}
where $\bvec_1(\cdot)$ shifts and scales, $\bvec_2(\cdot)$ rotates around the origin, and $\bvec_3(\cdot)$ reflects around the horizontal axis. Figure \ref{fig:warping_viz} in the Supplementary Material illustrates the effect of the homogenizing function $\bvec_0(\cdot)$ on points in $D$. 

Denote $\tilde\svec_k \equiv \fvec(\svec_k), \tilde\svec_l \equiv \fvec(\svec_l)$, and $\tilde\svec_m \equiv \fvec(\svec_m)$. The shifting and scaling is done through the function $ \bvec_1(\svec) \equiv \frac{1}{\norm{\tilde\svec_l - \tilde\svec_k}} (\svec - \tilde\svec_k); \quad \svec \in D. $
Denote the scaled and shifted domain as $D_1$, where $D_1 \equiv \{\bvec_1(\svec): \svec \in D\}$. Note that the distance between $\tilde\svec_l$ and $\tilde\svec_k$ is fixed to be 1 in $D_1$.  The operation that rotates $\bvec_1(\tilde\svec_l)$ to the point $(1, 0)'$ is given by
$$ \bvec_2(\svec) \equiv \begin{pmatrix}
\cos \psi_l  &  \sin \psi_l \\
-\sin \psi_l &  \cos \psi_l  \\
\end{pmatrix}\svec  ; \quad \svec \in D_1, $$
where \Copy{rev3iip9}{$\psi_l = \text{atan2}(b_{1,2}(\tilde\svec_l), b_{1,1}(\tilde\svec_l))$ is the angle of $\bvec_1(\tilde\svec_l)$, and $b_{1,i}(\cdot)$ refers to the $i^{\text{th}}$ element of $\bvec_1(\cdot)$.} Denote the scaled, shifted, and rotated domain as $D_2$, where $D_2 \equiv \{\bvec_2(\svec): \svec \in D_1\}$. Finally, the reflection operation that ensures that $b_{0,2}(\tilde\svec_m) > 0$ is given by
$$ \bvec_3(\svec) \equiv \begin{pmatrix} 1 &  0 \\
0 &  g_m 
\end{pmatrix}\svec  ; \quad \svec \in D_2, $$
where $g_m \equiv \sign(\bvec_2 \circ \bvec_1(\tilde\svec_m))$, equal to $-1$ if a reflection around the horizontal axis is needed, and equal to $+1$ otherwise. The fixed frame of reference is defined to be the domain $D_3 \equiv \{\bvec_3(\svec): \svec \in D_2\}$.

Fixing the frame of reference can be useful when, for example, one is bootstrapping to do uncertainty quantification of the warped locations, since these warped locations are non-identifiable otherwise. Importantly, we have the following result when the covariance functions in the deformed space are solely functions of (scaled) distances.

\begin{theorem}\label{thm:homogenizing}
	Assume that the cross-covariance functions on the warped domain, $\tilde{C}_{ij,D}^{o}(\hvec; a_{ij})$, where $\{a_{ij} \}$ are scale parameters, are solely functions of $a_{ij}\|\hvec\|, \hvec \in \mathbb{R}^2, a_{ij} > 0$. Consider two cross-covariance matrix functions $\Cmat_G^{(1)}(\cdot, \cdot)$ and $\Cmat_G^{(2)}(\cdot, \cdot)$, respectively, where $C_{ij,G}^{(r)}(\svec,\uvec) \equiv \tilde C_{ij,D}^{o}(\norm{\fvec^{(r)}(\svec) - \fvec^{(r)}(\uvec)}; a_{ij}^{(r)})$, for  $r = 1,2$, and $\svec,\uvec \in G$. If $\Cmat_G^{(1)}(\cdot, \cdot) = \Cmat_G^{(2)}(\cdot, \cdot)$, then $\bvec_0 \circ \fvec^{(1)}(\cdot) = \bvec_0 \circ \fvec^{(2)}(\cdot)$, where $\bvec_0(\cdot)$ is given by \eqref{eq:b_func}, and $a_{ij}^{(1)} \norm{\fvec^{(1)}(\svec_l) - \fvec^{(1)}(\svec_k)} = a_{ij}^{(2)} \norm{\fvec^{(2)}(\svec_l) - \fvec^{(2)}(\svec_k)}$. Conversely, if $\bvec_0 \circ \fvec^{(1)}(\cdot) = \bvec_0 \circ \fvec^{(2)}(\cdot)$, and $a_{ij}^{(1)} \norm{\fvec^{(1)}(\svec_l) - \fvec^{(1)}(\svec_k)} = a_{ij}^{(2)} \norm{\fvec^{(2)}(\svec_l) - \fvec^{(2)}(\svec_k)}$, for $i,j = 1,\dots,p$, then $\Cmat_G^{(1)}(\cdot, \cdot) = \Cmat_G^{(2)}(\cdot, \cdot)$.
	
\end{theorem}

\noindent \textit{Proof}: See Section \ref{sec:thm:homogenizing_proof} of the Supplementary Material.

Theorem \ref{thm:homogenizing} shows that, after homogenization using \eqref{eq:b_func}, locations warped using functions that yield the same cross-covariance matrix functions on the geographic domain, must coincide. This result can be used to obtain a visual appreciation of the uncertainty in the estimated warping function when bootstrapping the warping parameters: Informally, after homogenization, two covariance functions that are similar should yield points that are in close proximity to one another, and vice versa. We use such a visual diagnostic in our simulation study in Section \ref{sec:simstudies_sym}.

Theorem \ref{thm:homogenizing} also reveals that there is a one-to-one correspondence between the cross-covariance matrix function on the geographic domain and the scale parameters in the warped domain after homogenization. Specifically, $\tilde{a}_{ij} = a_{ij}^{(1)} \norm{\fvec^{(1)}(\svec) - \fvec^{(1)}(\uvec)} = a_{ij}^{(2)} \norm{\fvec^{(2)}(\svec) - \fvec^{(2)}(\uvec)}$, for $i,j = 1,\dots,p$, if and only if $\Cmat_G^{(1)}(\svec, \uvec) = \Cmat_G^{(2)}(\svec, \uvec)$, for all $\svec, \uvec \in G$. This leads to the following corollary, which shows that consistent inference of a transformation of the different process' scale parameters in the warped domain can be made after homogenizing the warpings to a fixed frame of reference. This can be useful for validating our methods when the true warping function is known, as it is in the simulation study presented in Section \ref{sec:simstudies_sym}.

\begin{corollary}\label{coll:scale_param}
	\Copy{rev3iip7}{Assume the conditions of Theorem \ref{thm:homogenizing}, and define $\tilde{a}_{ij} \equiv a_{ij} \norm{\fvec(\svec_l) - \fvec(\svec_k)},$ for $i, j = 1,\dots,p$. Then, the set comprising the homogenized warping function and transformed scale parameters, $\{ \bvec_0 \circ \fvec(\cdot), \{ \tilde a_{ij} \} \}$, is identifiable. That is, two sets of parameters $\{ \bvec_0 \circ \fvec^{(1)}(\cdot), \{ \tilde a^{(1)}_{ij} \} \}$ and $\{ \bvec_0 \circ \fvec^{(2)}(\cdot), \{ \tilde a^{(2)}_{ij} \} \}$, where $\tilde a^{(r)}_{ij} = a^{(r)}_{ij} \norm{\fvec^{(r)}(\svec_l) - \fvec^{(r)}(\svec_k)}, r = 1,2$, yield the same log restricted likelihood function $\mathcal{L}^{(1)}(\cdot; \cdot)$ and $\mathcal{L}^{(2)}(\cdot; \cdot)$ if and only if they are identical. That is, $\{ \bvec_0 \circ \fvec(\cdot), \{ \tilde a_{ij} \} \}$, is identifiable in the sense of \cite{kadane1974role}.}
\end{corollary}

\noindent \textit{Proof}: See Section \ref{sec:thm:homogenizing_proof} of the Supplementary Material.

Fixing the frame of reference allows us to do uncertainty quantification on any warping-function parameters and transformed scale parameters. While under certain conditions, REML estimators are asymptotically Gaussian \citep{cressie1996asymptotics}, we are not aware of an analytical form of the asymptotic distribution of the REML estimators for a nonstationary covariance model constructed through deformation. Hence, we use bootstrapping to make inference on these parameters. Bootstrapping with spatial data needs to be done with care, since the data are correlated; see \cite{solow1985bootstrapping} and \cite{olea2011generalized} for more discussion. A bootstrapping algorithm for quantifying the uncertainties of the parameters in model \eqref{eq:gau_model} is shown in Algorithm \ref{alg:bootstrap} in Section \ref{sec:additional_results} of the Supplementary Material \citep{olea2011generalized}. We use Algorithm \ref{alg:bootstrap} for visualizing uncertainties on warped locations, and we use it for uncertainty quantification of parameter estimates in the simulation study of Section \ref{sec:simstudies_sym}.

\section{Data Illustrations ~}\label{sec:illustrations}

In this section, we show the potential benefit of using multivariate DCSMs over conventional ones through two illustrations. In Section \ref{sec:simstudies_sym}, we show results from a study using data simulated from a symmetric nonstationary bivariate-covariance model. In Section \ref{sec:applications_ocean}, we show results from a study using North Atlantic Ocean temperatures at two different depths. Section \ref{sec:additional_illustrations} of the Supplementary Material contains additional data illustrations, using data simulated from an asymmetric nonstationary bivariate covariance model, from models with misspecified warping functions, and from a trivariate covariance model. It also contains an experiment using real maximum-and-minimum-temperature data in the United States. Code and data for reproducing the results from all our data illustrations are available from \url{https://github.com/quanvu17/deepspat_multivar}.

\subsection{Simulated Symmetric Nonstationary Data ~}\label{sec:simstudies_sym}

We first demonstrate the use of multivariate DCSMs on data simulated using a symmetric nonstationary bivariate covariance model. We simulated the bivariate data from a Gaussian multivariate DCSM, $\tilde \Yvec(\cdot)$, with constant mean (i.e., $q = 1$ and $\xvec(\cdot) = x_1(\cdot) = 1$ in \eqref{eq:y_model}, so that there are two intercepts, $\beta_{11}$ and $\beta_{21}$, that need to be estimated). The data were simulated on an equally spaced 101 $\times$ 101 grid of the geographic domain, $G \equiv [-0.5, 0.5] \times [-0.5, 0.5]$. The warping function we used was a composition of axial warping units, followed by a single-resolution radial basis function unit, followed by a M{\"o}bius transformation unit; see \cite{azm2019deep} for a detailed description of these warping units. On the warped domain, we modeled the covariances using a stationary, isotropic, multivariate parsimonious Mat{\'e}rn model. We randomly sampled 1000 locations from the grid and used these as measurement locations.

\begin{figure}
	\centering
	
	\includegraphics[width=0.8\textwidth]{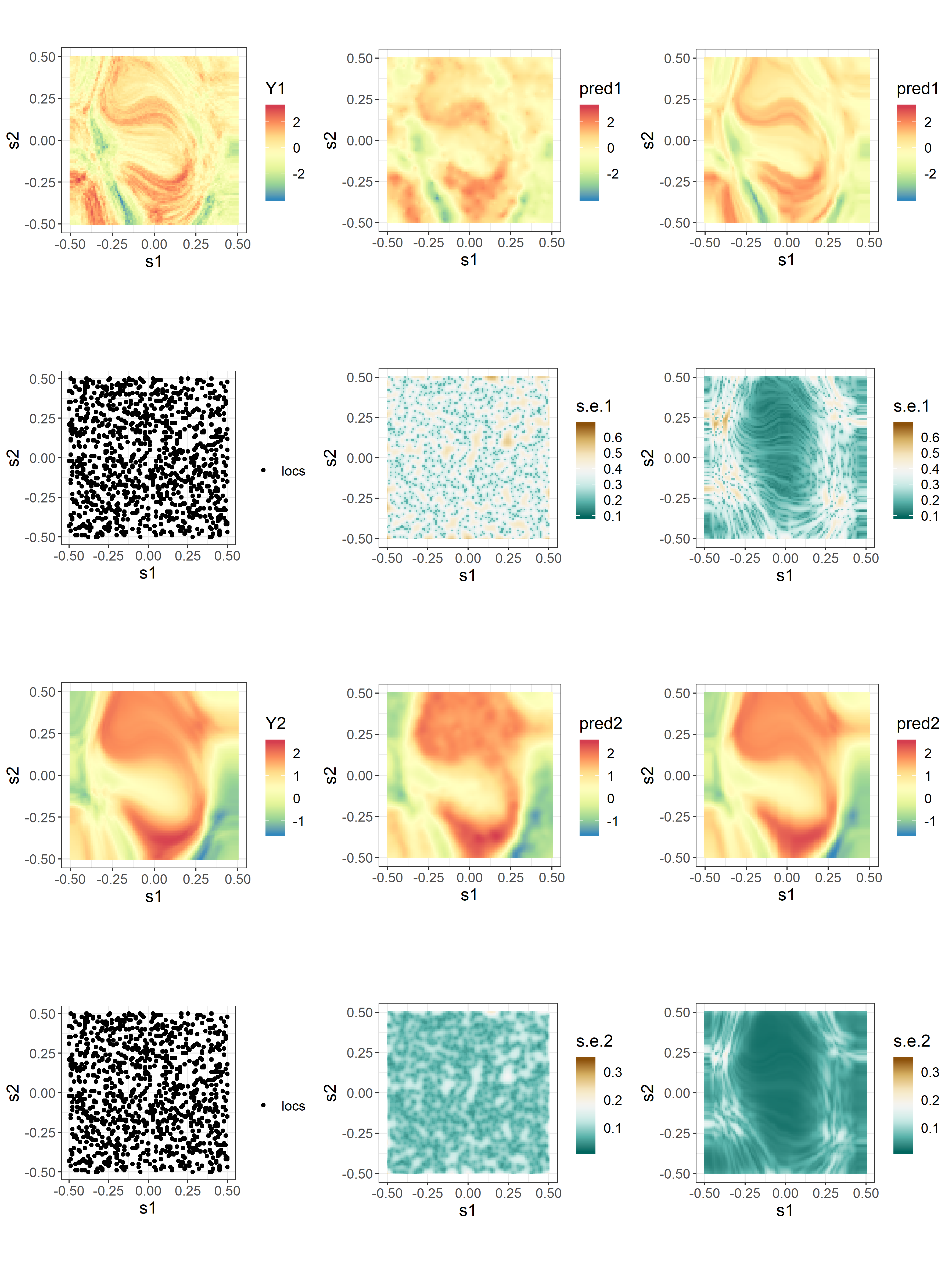}
	
	\caption{
		\footnotesize{
			Comparison of predictions and prediction standard errors when using a bivariate stationary parsimonious Mat{\'e}rn model (Model 4.1.1) and a bivariate DCSM (Model 4.1.2) in the study of Section \ref{sec:simstudies_sym}, where measurement locations were randomly sampled from $G$. 
			First row: The process $\tilde Y_1(\cdot)$ (left panel), the predictions obtained using Model 4.1.1 (center panel), and Model 4.1.2 (right panel). Second row: Locations of the measurement of $\tilde Y_1(\cdot)$ (left panel), the prediction standard errors obtained when using Model 4.1.1 (center panel) and Model 4.1.2 (right panel). Third and fourth rows: Analogous to the first and second rows, respectively, for the process $\tilde Y_2(\cdot)$.}
	}
	\label{fig:sim_study}
\end{figure}

We compared the predictions of the stationary parsimonious Mat{\'e}rn model (Model 4.1.1) to those of the multivariate DCSM (Model 4.1.2), in order to gauge the loss in prediction performance when the nonstationarity arising from the warping is ignored. After fitting Model 4.1.1 and Model 4.1.2 to the observations at the 1000 locations, we computed the predictions and prediction standard errors of the latent processes on the 101 $\times$ 101 grid. Figure \ref{fig:sim_study} shows the true simulated fields, the predictions, and the prediction standard errors, from both models. As was observed in the univariate case \citep{azm2019deep}, we see that the DCSM can predict sharp features in the spatial fields, while the stationary parsimonious Mat{\'e}rn model smooths out such features. Further, while the stationary parsimonious Mat{\'e}rn model produces prediction standard errors that are mostly unrelated to the process behavior (due to the stationarity assumption), the DCSM produces prediction standard errors that are highly reflective of the processes' local anisotropies and scales. These visualizations illustrate the advantages of using a multivariate DCSM over a stationary multivariate model when the underlying processes are highly nonstationary.

To compare the predictive performance of the two models quantitatively, we calculated the predictive performance at the remaining 9201 locations using two commonly used scoring rules, namely the root-mean-square prediction error (RMSPE) and the continuous ranked probability score (CRPS) \citep{gneiting2007strictly}. We also repeated the procedure of random sampling 1000 locations and accessing the predictive performance, for 30 times. Table \ref{tbl:crossvalid_sim} summarizes the results with averages from the 30 cross-validation studies, and also gives the average Akaike information criterion (AIC) from these studies. Figure \ref{fig:sim_symm_boxplot} in the Supplementary Material shows the boxplots of the RMSPE and CRPS for both models across the 30 simulations. From the table and figure, it is clear that there is a large improvement in RMSPE and CRPS when using the DCSM (Model 4.1.2) over the stationary parsimonious Mat{\'e}rn model (Model 4.1.1). This improvement was expected since the data were generated from the highly nonstationary process. Use of the more sophisticated nonstationary model comes at some computational cost: It took 1545.4 seconds on average to fit Model 4.1.2, almost twice as long as the 823.5 seconds it took to fit Model 4.1.1.

\begin{table}
	\centering
	\caption{Average hold-out-validation results, AIC, and the time required to fit, for the simulation study in Section \ref{sec:simstudies_sym}, where the measurement locations are randomly sampled 30 times from $G$.}
	\label{tbl:crossvalid_sim}
	\bgroup
	\def\arraystretch{1}
	\begin{tabular}{ |c|c|c|c|c|c|c| }
		\hline
		& \multicolumn{2}{c|}{$\tilde Y_1(\cdot)$} & \multicolumn{2}{c|}{$\tilde Y_2(\cdot)$} & & \\
		\cline{2-5}
		& RMSPE & CRPS & RMSPE & CRPS & AIC & Time (s) \\
		\hline
		Model 4.1.1 & 0.404 &	0.221 & 0.089 & 0.048 & 887.0 & 823.5 \\
		\hline
		Model 4.1.2 & 0.358 &	0.196 & 0.072 & 0.037 & 257.0 & 1545.4 \\
		\hline
	\end{tabular}
	\egroup
\end{table}

\begin{figure}[t!]
	\centering
	\includegraphics[width=0.6\textwidth]{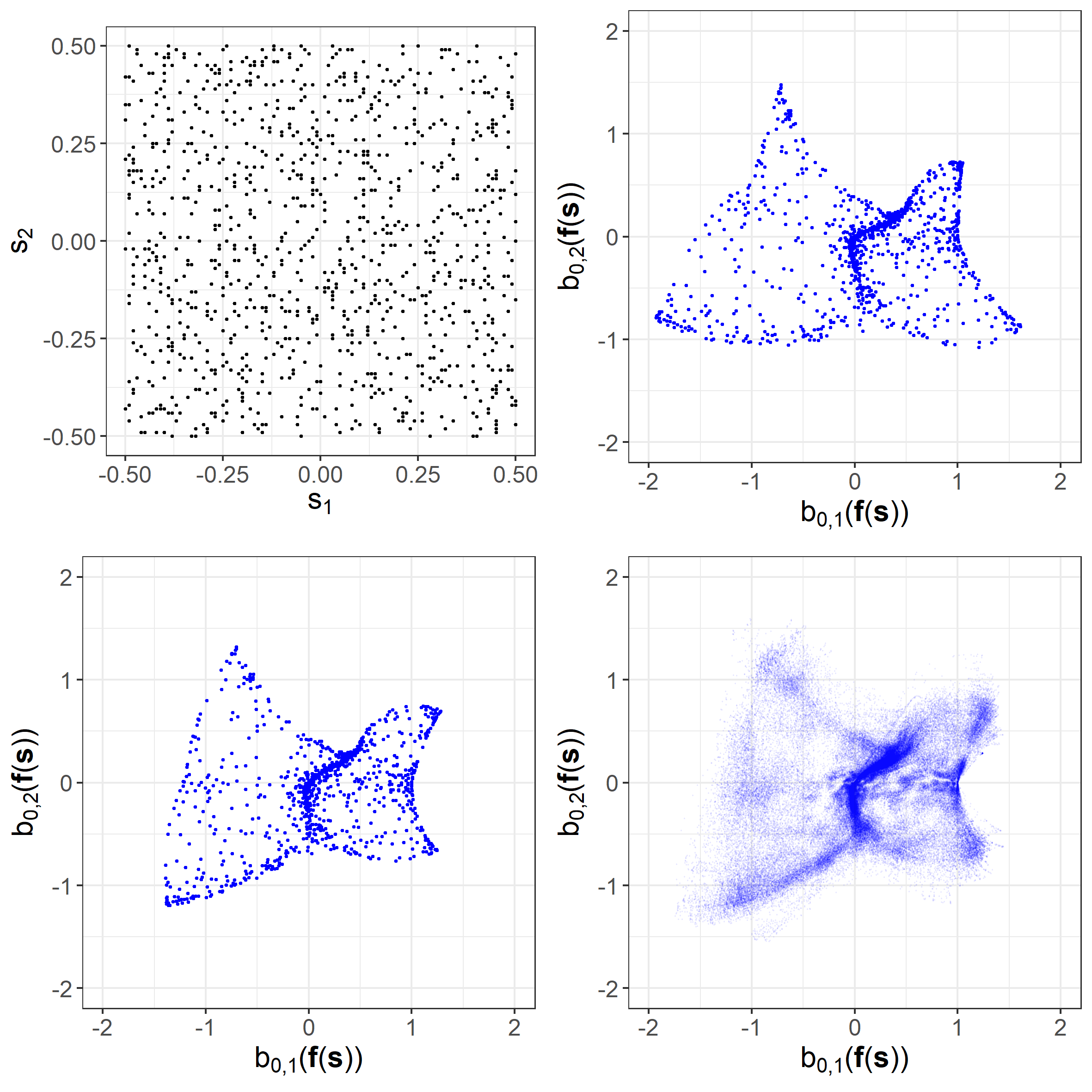}
	\caption{
		\footnotesize{Measurement locations on the original domain $G$ and the warped domain after homogenization, $D_3$. Top row: Measurement locations on the original domain (left panel); true warped measurement locations (right panel). Bottom row: REML estimate of the warped measurement locations (left panel); bootstrap distribution of the warped measurement locations (right panel; for visualization purposes only 100 bootstrap samples are shown). }
	}
	\label{fig:warped_loc}
\end{figure}

We next used a bootstrap to examine the ability of the DCSM to recover the true parameters. We bootstrapped in a fixed frame of reference (via the homogenizing function) with 1000 bootstrap samples to quantify uncertainties on the model parameters using the method outlined in Section \ref{sec:frame_of_ref}. Figure \ref{fig:warped_loc} shows the measurement locations in $G$, the measurement locations under the true warping function and homogenization, the measurement locations under the estimated warping function and homogenization, and the bootstrap samples of the warped locations. We see that the estimated warped locations and the bootstrap samples of the warped locations are similar to the warped locations under the true warping function. Specifically, important features, such as the contraction in the middle part of the domain, are recovered. Table \ref{tbl:est_par} in the Supplementary Material lists the true cross-covariance matrix function parameters along with their estimates and their 95\% bootstrap confidence intervals. The REML estimates are relatively close to the true value, and all the 95\% bootstrap confidence intervals of the model parameters contain the true values.

We also considered the case where the data are missing in a block, which is shown in Section \ref{sec:simstudies_sym_block} in the Supplementary Material.

\subsection{Modeling Temperatures in the North Atlantic Ocean at Two Different Depths ~}\label{sec:applications_ocean}

\begin{figure}
	\centering
	
	\includegraphics[width=1\textwidth]{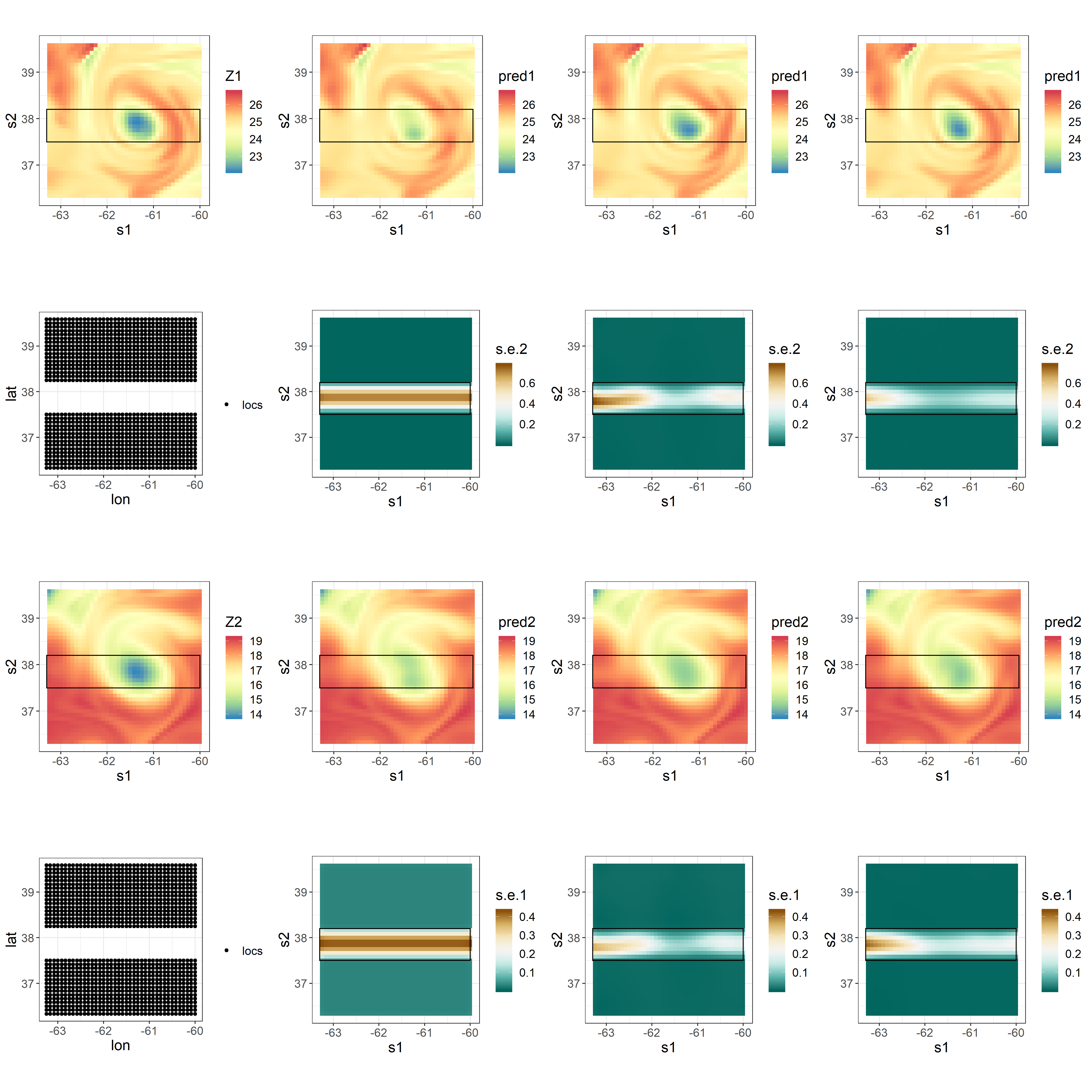}
	
	\caption{
		\footnotesize{
			Comparison of predictions when using a symmetric stationary parsimonious Mat{\'e}rn model (Model 4.2.1), a bivariate symmetric DCSM (Model 4.2.2), and a bivariate asymmetric DCSM (Model 4.2.3).
			First row: Original temperature observations at depth 0.5 meters, $Z_1$ (first panel); predictions obtained using Model 4.2.1 (second panel), Model 4.2.2 (third panel), and Model 4.2.3 (fourth panel).
			Second row: Locations of the retained measurements (first panel); prediction standard errors obtained when using Model 4.2.1 (second panel), Model 4.2.2 (third panel), and Model 4.2.3 (fourth panel).
			Third and fourth rows: Analogous to the first and second rows, respectively, for $Z_2$, the temperature observations at depth 318.1 meters.
		}
	}
	\label{fig:ocean_gap_study}
\end{figure}

We next consider sea temperatures in the North Atlantic Ocean at two very different depths: 0.5 meters and 318.1 meters. The data were obtained from the Copernicus Marine Environment Monitoring Service (CMEMS)\footnotetext{$^1$\url{http://marine.copernicus.eu/services-portfolio/access-to-products/?option=com_csw&view=details&product_id=GLOBAL_ANALYSIS_FORECAST_PHY_001_024}}\footnotemark{$^1$}. We analyzed temperatures on 1 July 2018 between 36.3$^{\circ}$N--39.6$^{\circ}$N and 60.0$^{\circ}$W--63.3$^{\circ}$W, with 1600 measurements in this region whose locations were on a $40 \times 40$ grid. Panels (1,1) and (3,1) in Figure \ref{fig:ocean_gap_study} show the temperatures at the two depths, where we can see that there is a small amount of misalignment in the temperature processes, suggesting that the modeling of cross-covariance asymmetry may be important.

We considered the following models.
\begin{itemize}
	\item Model 4.2.1: A bivariate stationary, symmetric, parsimonious Mat{\'e}rn model with only an intercept in the trend.
	\item Model 4.2.2: A bivariate symmetric DCSM, where the warping function $\fvec(\cdot)$ is a composition of axial warping units, a single-resolution radial basis function unit, and a M{\"o}bius transformation unit, and with Model 4.2.1 on the warped domain.
	\item Model 4.2.3: A bivariate asymmetric DCSM, where the aligning function $\gvec_2(\cdot)$ is an affine transformation (as described in Proposition \ref{propo:marginal_stat_model}), and the warping function is as in Model 4.2.2, and with Model 4.2.1 on the warped domain.
\end{itemize}

The predictive performance of these three models was first examined using a five-fold cross-validation study, where we randomly divided the 1600 measurement locations into five groups. The results are summarized in Table \ref{tbl:crossvalid_ocean_temp} in the Supplementary Material. There, it can be seen that allowing for model nonstationarity and/or asymmetry does indeed result in improved predictions, but the observed improvement is not substantial.

We also considered the case where the data are missing in a block. Specifically, we assumed that we have all the measurements on the grid except those between 37.5$^{\circ}$N--38.2$^{\circ}$N. Figure \ref{fig:ocean_gap_study} shows the true fields and the predictions using Model 4.2.1, Model 4.2.2, and Model 4.2.3, while Table \ref{tbl:crossvalid_ocean_temp_gap} shows the diagnostic results when predicting the temperature at the missing locations. The improvement of the bivariate DCSMs over the stationary, symmetric Model 4.2.1 is evident in this case. Observe that the asymmetric version of the DCSM (Model 4.2.3) produces slightly better predictive diagnostics than the symmetric version (Model 4.2.2), illustrating the importance of having the ability to model asymmetry. Visualizations of the nonstationary and asymmetric structure of estimated cross-covariance matrix function are given in Figure \ref{fig:ocean_warping} and Figure \ref{fig:ocean_heatmap} in the Supplementary Material.

\begin{table}
	\centering
	\caption{
		Hold-out-validation results, AIC, and the time required to fit, for the ocean-temperature data at depths 0.5 meters and 318.1 meters for the study in Section \ref{sec:applications_ocean}, where data were missing in the white block shown in the second and fourth rows of Figure \ref{fig:ocean_gap_study}.
	}
	\label{tbl:crossvalid_ocean_temp_gap}
	\bgroup
	\def\arraystretch{1}
	\begin{tabular}{ |c|c|c|c|c|c|c| }
		\hline
		& \multicolumn{2}{c|}{$T_{0.5}$} & \multicolumn{2}{c|}{$T_{318.1}$} & & \\
		\cline{2-5}
		& RMSPE & CRPS & RMSPE & CRPS & AIC & Time (s) \\
		\hline
		Model 4.2.1 & 0.450 & 0.226 & 0.463 & 0.236 & -5274.7 & 1311.1 \\
		\hline
		Model 4.2.2 & 0.234 & 0.136 & 0.301 & 0.176 & -6891.7 & 2491.0 \\
		\hline
		Model 4.2.3 & 0.228 & 0.126 & 0.301 & 0.166 & -7892.1 & 4343.2 \\
		\hline
	\end{tabular}
	\egroup
\end{table}

\section{Conclusion ~}\label{sec:conclusion}

In this article, we introduce a new class of cross-covariance matrix functions that are valid by construction and that are able to capture both nonstationarity and asymmetry. Specifically, through $p$ warping functions, each of which is constructed as a composition of elementary injective warping units, we model $p$-variate spatial processes that have nonstationary and asymmetric covariances on the geographic domain. These are modeled in terms of processes with stationary, symmetric, and possibly isotropic covariances on a warped domain. We also consider a special case where the same warping function is used for all $p$ processes, resulting in a symmetric cross-covariance matrix function on the geographic domain. We show the benefit of using the multivariate DCSMs over classical stationary models, such as the multivariate parsimonious Mat{\'e}rn model, through illustrations based on simulated data and real-world data.

There are a number of avenues that can be considered for future development of the proposed models. First, the models we propose do not consider nonstationarity in the variance parameters or in the cross-correlation parameters. Spatially varying variance parameters and cross-correlation parameters can be introduced as in \cite{kleiber2012nonstationary} and \cite{messick2017multivariate}. Second, in this article we only make use of the parsimonious multivariate Mat{\'e}rn model on the warped domain, but indeed any model could be used (e.g., one based on the cross-variogram).  \Copy{rev3iip11}{Third, we present data examples for two-dimensional space ($d=2$), but our model can also be used in higher-dimensional space. The challenge is to find warping functions that are injective in a higher-dimensional domain.} \Copy{compute_limit}{Further, since the computational complexity of evaluating the likelihood function in the model is $O(N^3)$ for a data set of size $N$, the model needs to be modified in a large-spatial-data setting. Specifically, to deal with very large nonstationary spatial data sets, one would need to extend the model in such a way that it has a scalable structure on the deformed space. Several models that can deal with very large spatial data sets are summarized in \cite{heaton2019case}.} \Copy{architecture_limit}{Finally, when using our multivariate spatial models, several decisions need to be made on the architecture (e.g., the number of layers of warping units, the order of the warping units, etc.), and further work needs to be carried out to determine how these decisions affect predictive performance.}

In conclusion, we show that multivariate DCSMs are easy to construct and then fit from simple injective warping functions. We also show that they can provide superior predictive performance compared to conventional stationary models, particularly when data are missing over large regions.

\vskip 14pt
\noindent {\large\bf Supplementary Material}

Section \ref{sec:appendix_proof} contains the proofs of Propositions 1--4, Theorem 1, and Corollary 1; Section \ref{sec:predictions} gives the log restricted likelihood and prediction formulas; Section \ref{sec:additional_results} contains additional tables and figures; and Section \ref{sec:additional_illustrations} contains additional data illustrations.

\par

\vskip 14pt
\noindent {\large\bf Acknowledgements}

Quan Vu was supported by a University Postgraduate Award from the University of Wollongong, Australia. Andrew Zammit-Mangion's research was supported by an Australian Research Council Discovery Early Career Research Award (DECRA) DE180100203 and by Discovery Project DP190100180. Noel Cressie's research was supported by Australian Research Council Discovery Projects DP150104576 and DP190100180, and by NSF grant SES-1132031 funded through the NSF-Census Research Network (NCRN) program. Cressie's and Zammit-Mangion's research was also supported by NASA ROSES grant 17-OCO2-17-0012.  The authors are grateful to the Associate Editor and three referees whose suggestions led to a number of improvements in the manuscript. They would also like to thank Matt Moores for discussions related to this research.
\par

\newpage

\appendix

\begin{center}
\Large \bf Supplementary Material
\end{center}

\setcounter{section}{0}
\setcounter{equation}{0}
\setcounter{table}{0}
\setcounter{figure}{0}
\def\theequation{S\arabic{section}.\arabic{equation}}
\def\thesection{S\arabic{section}}

\def\thefigure{S\arabic{figure}}
\def\thetable{S\arabic{table}}

\section{Proofs ~}\label{sec:appendix_proof}

\subsection{Proof of Proposition \ref{propo:symm} ~}\label{sec:propo:symm_proof}
Since $\Cvec^{o}_{D}(\cdot)$ is symmetric, $C_{ij,D}^o(\cdot) = C_{ji,D}^o(\cdot)$, $i, j = 1, \dots, p$. Now consider any two locations $\svec,\uvec \in G$. The covariance between $Y_i(\svec)$ and $Y_j(\uvec)$ is given by
\begin{equation*}
	C_{ij, G}(\svec, \uvec) = C^{o}_{ij, D}({\fvec(\svec) - \fvec(\uvec)}) 
	= C^{o}_{ji, D}({\fvec(\svec) - \fvec(\uvec)}) 
	= C_{ji, G}(\svec, \uvec),
\end{equation*}
for $i, j = 1, \dots, p$, and therefore $\Cmat_G(\cdot \, , \cdot)$ is symmetric. 

\subsection{Proof of Proposition \ref{propo:asymmetry} ~}\label{sec:propo:asymmetry_proof}
Consider any two locations $\svec, \uvec$ in $G$. The covariance between $Y_i(\svec)$ and $Y_j(\uvec)$, $i, j = 1, \dots, p$, is
\begin{align*}
	\cov(Y_i(\svec), Y_j(\uvec)) =  C_{ij, G}(\svec, \uvec) &= C^{o}_{ij, D}({\fvec_i(\svec) - \fvec_j(\uvec)}),
\end{align*}
while the covariance between $Y_j(\svec)$ and $Y_i(\uvec)$ is
\begin{align*}
	\cov(Y_j(\svec), Y_i(\uvec)) = C_{ji, G}(\svec, \uvec) &= C^{o}_{ji, D}({\fvec_j(\svec) - \fvec_i(\uvec)}) \\
	&= C^{o}_{ij, D}({\fvec_j(\svec) - \fvec_i(\uvec)}),
\end{align*}
since $\Cvec^{o}_D(\cdot)$ is symmetric. Now, for $i \neq j$, there are many examples where $\fvec_i(\svec) - \fvec_j(\uvec) \neq \fvec_j(\svec) - \fvec_i(\uvec)$, and hence $\cov(Y_i(\svec), Y_j(\uvec)) \neq \cov(Y_j(\svec), Y_i(\uvec))$ for $i \neq j$. That is, the cross-covariance matrix function $\Cvec_G(\cdot, \cdot)$ constructed through \eqref{eq:cov_model_asym_general} is not necessarily symmetric. 

\subsection{Proof of Proposition \ref{propo:valid} ~}\label{sec:propo:valid_proof}
By \eqref{eq:cov_model_asym_general}, we have that for $k = 1,\dots,n_i, ~ l = 1,\dots,n_j$, and $i,j = 1,\dots, p$, $C_{ij, G}(\svec_{ik}, \svec_{jl}) = C^o_{ij, D}({\fvec_i(\svec_{ik}) - \fvec_j(\svec_{jl})})$. Therefore, for $i,j = 1,\dots,p$, we have that  $\Sigmavec_{ij, G} = (C^o_{ij, D}({\fvec_i(\svec_{ik}) - \fvec_j(\svec_{jl})}): k = 1,\dots,n_i, ~ l = 1,\dots,n_j)$. Since $\Cvec^o_{D}(\cdot)$ is valid, we have that for any $\avec \in \mathbb{R}^{N}, \avec \neq \zerovec$,
\( \avec' (C^o_{ij, D}(\fvec_{i}(\svec_{ik}) - \fvec_{j}(\svec_{jl})): k = 1,\dots,n_i; ~ l = 1,\dots,n_j; ~ i,j = 1,\dots, p) \avec \geq 0, \)
and hence $\avec'\Sigmavec_G \avec \ge 0$. That is, $\Sigmavec_G$ is nonnegative-definite.

\subsection{Proof of Proposition \ref{propo:marginal_stat_model} ~}\label{sec:propo:marginal_stat_model_proof}
Since $\fvec(\cdot)$ is the identity map,
\begin{align*}
	C_{ij, G}(\svec, \uvec) &= C^{o}_{ij, D}(\gvec_i(\svec) -\gvec_j(\uvec)); \quad i,j \neq k, \\
	C_{kk, G}(\svec, \uvec) &= C^{o}_{kk, D}(\gvec_k(\svec) -\gvec_k(\uvec)), \\
	C_{ik, G}(\svec, \uvec) &= C^{o}_{ik, D}(\gvec_i(\svec) -\gvec_k(\uvec)); \quad i \neq k,
\end{align*}
where $k \in \{2,\dots,p\} $. Write the affine transformation as $\gvec_k(\svec) = \tilde{\mathbf{A}} \svec + \tilde{\mathbf{d}}$, where $\tilde{\mathbf{A}}$ is a $d \times d$ matrix that is not generally equal to the identity matrix. Then,
\begin{align*}
	C_{ij, G}(\svec, \uvec) &= C^{o}_{ij, D}(\svec - \uvec); \quad i,j \neq k, \\
	C_{kk, G}(\svec, \uvec) &= C^{o}_{kk, D}(\tilde{\mathbf{A}} \svec - \tilde{\mathbf{A}} \uvec) = C^{o}_{kk, D}(\tilde{\mathbf{A}} (\svec - \uvec)),\\
	C_{ik, G}(\svec, \uvec) &= C^{o}_{ik, D}(\svec - \tilde{\mathbf{A}} \uvec - \tilde{\mathbf{d}}); \quad i \neq k.
\end{align*}
As $\Cvec^o_{D}(\cdot)$ is stationary, it follows that $C_{ij, G}(\svec, \uvec), i,j \neq k$ and $C_{kk, G}(\svec, \uvec)$ are stationary, while $C_{ik, G}(\svec, \uvec), i \neq k$, is nonstationary unless $\tilde{\mathbf{A}}$ is the identity matrix. That is, the cross-covariance function $C_{ik,G}(\cdot, \cdot)$ is not necessarily stationary.

\subsection{Proof of Theorem \ref{thm:homogenizing} and Corollary \ref{coll:scale_param} ~}\label{sec:thm:homogenizing_proof}

\textbf{Proof of Theorem \ref{thm:homogenizing}: \\ }
By assumption,  $\Cmat_G^{(1)}(\svec, \uvec) = \Cmat_G^{(2)}(\svec, \uvec)$, for all $\svec,\uvec \in G$, and hence by definition, $\tilde C_{ij,D}^{o}(\norm{\fvec^{(1)}(\svec) - \fvec^{(1)}(\uvec)}; a_{ij}^{(1)}) = \tilde C_{ij,D}^{o}(\norm{\fvec^{(2)}(\svec) - \fvec^{(2)}(\uvec)}; a_{ij}^{(2)})$, for all $\svec,\uvec \in G$ and $i,j = 1,\dots,p$. That is, for each $\svec, \uvec \in G$ and $i,j=1,\dots,p$, we have that $a_{ij}^{(1)} \norm{\fvec^{(1)}(\svec) - \fvec^{(1)}(\uvec)} = a_{ij}^{(2)} \norm{\fvec^{(2)}(\svec) - \fvec^{(2)}(\uvec)}$, and hence $a_{ij}^{(1)} \norm{\fvec^{(1)}(\svec_l) - \fvec^{(1)}(\svec_k)} = a_{ij}^{(2)} \norm{\fvec^{(2)}(\svec_l) - \fvec^{(2)}(\svec_k)}$, for $\svec_l$ and $\svec_k$ two distinct points in $G$.

Hence, from the definition of $\bvec_1(\cdot)$ in Section \ref{sec:frame_of_ref},
\[ \norm{\bvec_1 \circ \fvec^{(1)}(\svec) - \bvec_1 \circ \fvec^{(1)}(\uvec)} = \norm{\bvec_1 \circ \fvec^{(2)}(\svec) - \bvec_1 \circ \fvec^{(2)}(\uvec)}, \]
for all $\svec,\uvec \in G$. Since $\bvec_2(\cdot)$ and $\bvec_3(\cdot)$ are distance-preserving transformations, we then have that
\[ \norm{\bvec_0 \circ \fvec^{(1)}(\svec) - \bvec_0 \circ \fvec^{(1)}(\uvec)} = \norm{\bvec_0 \circ \fvec^{(2)}(\svec) - \bvec_0 \circ \fvec^{(2)}(\uvec)}. \]

\noindent Recall that three locations $\svec_k$, $\svec_l$, and $\svec_m$ are chosen in $G \subset \mathbb{R}^2$ such that $\fvec^{(r)}(\svec_k)$, $\fvec^{(r)}(\svec_l)$, and $\fvec^{(r)}(\svec_m)$ are not colinear; and $\bvec_0\circ\fvec^{(r)}(\svec_k) = (0,0)'$,  $\bvec_0\circ\fvec^{(r)}(\svec_l) = (1,0)'$, and $b_{0,2}\circ\fvec^{(r)}(\svec_m) > 0$, for $r = 1,2$. Now, we have that for any two points $\svec, \uvec \in G$, the distance $\norm{\bvec_0 \circ \fvec^{(r)}(\svec) - \bvec_0 \circ \fvec^{(r)}(\uvec)}$ does not depend on $r$. Because the two points $\bvec_0 \circ \fvec^{(r)}(\svec_k)$ and $\bvec_0 \circ \fvec^{(r)}(\svec_l)$ are fixed in $D_3$, it follows that the distances $\norm{\bvec_0 \circ \fvec^{(r)}(\svec_m) - \bvec_0 \circ \fvec^{(r)}(\svec_k)}$ and $\norm{\bvec_0 \circ \fvec^{(r)}(\svec_m) - \bvec_0 \circ \fvec^{(r)}(\svec_l)}$ are also fixed. In $\mathbb{R}^2$, when the distances from a point $\bvec_0 \circ \fvec^{(r)}(\svec_m)$ to two fixed points $(0,0)'$ and $(1,0)'$ are constant, there exist two possible points $\svec_m$, assuming $\fvec^{(r)}(\svec_m)$ is not colinear with $\fvec^{(r)}(\svec_k)$ and $\fvec^{(r)}(\svec_l)$: one where $b_{0,2} \circ \fvec^{(r)}(\svec_m) > 0$ and one where $b_{0,2} \circ \fvec^{(r)}(\svec_m) < 0$. However, as we constrain $b_{0,2} \circ \fvec^{(r)}(\svec_m) > 0$, $\bvec_0 \circ \fvec^{(r)}(\svec_m)$ is unique. Since we have three fixed points $\bvec_0 \circ \fvec^{(r)}(\svec_k)$,  $\bvec_0 \circ \fvec^{(r)}(\svec_l)$, and  $\bvec_0 \circ \fvec^{(r)}(\svec_m)$, and a fixed set of distances $\norm{\bvec_0 \circ \fvec^{(r)}(\svec) - \bvec_0 \circ \fvec^{(r)}(\uvec)}$ for any two points $\svec, \uvec$, then $\bvec_0 \circ \fvec^{(1)}(\svec) = \bvec_0 \circ \fvec^{(2)}(\svec)$, for all $\svec \in G$.

For the converse part of the proof, assume that $\bvec_0 \circ \fvec^{(1)}(\svec) = \bvec_0 \circ \fvec^{(2)}(\svec)$, for all $\svec \in G$. Then,
\[ \norm{\bvec_1 \circ \fvec^{(1)}(\svec) - \bvec_1 \circ \fvec^{(1)}(\uvec)} = \norm{\bvec_1 \circ \fvec^{(2)}(\svec) - \bvec_1 \circ \fvec^{(2)}(\uvec)}, \]
for all $\svec,\uvec \in G$, and therefore
\[ \frac{\norm{\fvec^{(1)}(\svec) - \fvec^{(1)}(\uvec)}}{\norm{\fvec^{(1)}(\svec_l) - \fvec^{(1)}(\svec_k)}} = \frac{\norm{\fvec^{(2)}(\svec) - \fvec^{(2)}(\uvec)}}{\norm{\fvec^{(2)}(\svec_l) - \fvec^{(2)}(\svec_k)}}. \]
Because $a_{ij}^{(1)} \norm{\fvec^{(1)}(\svec_l) - \fvec^{(1)}(\svec_k)} = a_{ij}^{(2)} \norm{\fvec^{(2)}(\svec_l) - \fvec^{(2)}(\svec_k)}$, for $i,j = 1,\dots,p$, it follows that $a_{ij}^{(1)} \norm{\fvec^{(1)}(\svec) - \fvec^{(1)}(\uvec)} = a_{ij}^{(2)} \norm{\fvec^{(2)}(\svec) - \fvec^{(2)}(\uvec)}$, for all $\svec, \uvec \in G$ and $i,j = 1,\dots,p$. Therefore, $\Cmat_G^{(1)}(\svec, \uvec) = \Cmat_G^{(2)}(\svec, \uvec)$, for all $\svec,\uvec \in G$. \\

\noindent \textbf{Proof of Corollary \ref{coll:scale_param}: \\ }
Note that the Gaussian process model \eqref{eq:y_model} is fully specified by its mean function and covariance function. Hence, its finite-dimensional distributions are solely a function of the mean and covariance-function parameters. Therefore, the log restricted likelihood function in \eqref{eq:reml} where the mean-function parameters are profiled out, solely depends on covariance-function parameters and the data $\Zvec$. Now, suppose that two different sets each comprising a warping function and scale parameters, $\{ \fvec^{(1)}(\cdot), \{ a^{(1)}_{ij} \} \}$ and $\{ \fvec^{(2)}(\cdot), \{ a^{(2)}_{ij} \} \}$, yield the same log restricted likelihood function, for any set of measurement locations $\{ \svec_{ik}: k = 1,...,n_i; i = 1,...,p \} \subset G$. Then, this necessarily means that $\Cmat_G^{(1)}(\svec, \uvec) = \Cmat_G^{(2)}(\svec, \uvec)$, for all $\svec,\uvec \in G$ and, from the proof of Theorem \ref{thm:homogenizing} above, we see that this implies that $\bvec_0 \circ \fvec^{(1)}(\svec) = \bvec_0 \circ \fvec^{(2)}(\svec)$, for all $\svec \in G$, and that $a_{ij}^{(1)} \norm{\fvec^{(1)}(\svec_l) - \fvec^{(1)}(\svec_k)} = a_{ij}^{(2)} \norm{\fvec^{(2)}(\svec_l) - \fvec^{(2)}(\svec_k)}$, for $i,j = 1,\dots,p$. This in turn implies that the sets comprising the corresponding homogenized warping functions and transformed scale parameters, $\{ \bvec_0 \circ \fvec^{(1)}(\cdot), \{ \tilde a^{(1)}_{ij} \} \}$ and $\{ \bvec_0 \circ \fvec^{(2)}(\cdot), \{ \tilde a^{(2)}_{ij} \} \}$, where $\tilde a^{(r)}_{ij} = a^{(r)}_{ij} \norm{\fvec^{(r)}(\svec_l) - \fvec^{(r)}(\svec_k)}, r = 1,2$, are identical. Further, since Theorem \ref{thm:homogenizing} establishes an if-and-only-if relation, there is no other set, $\{\bvec_0 \circ \fvec^{(3)}(\cdot), \{ \tilde a_{ij}^{(3)} \} \}$ say, that yields the same log restricted likelihood function, for any set of measurement locations $\{ \svec_{ik}: k = 1,...,n_i; i = 1,...,p \} \subset G$. This is because such a set would correspond to a different covariance function, $\Cvec_G^{(3)}(\svec,\uvec)$ say, where $\Cvec_G^{(3)}(\svec,\uvec) \neq \Cvec_G^{(1)}(\svec,\uvec)$ for some $\svec, \uvec \in G$. Therefore, the set comprising a homogenized warping function and the transformed scale parameters, $\{ \bvec_0 \circ \fvec(\cdot), \{ \tilde a_{ij} \} \}$, where $\tilde{a}_{ij} \equiv a_{ij} \norm{\fvec(\svec_l) - \fvec(\svec_k)}$ for $i, j = 1,\dots,p$, has a unique log restricted likelihood function associated with it; this set is thus identifiable \citep[see][for more discussion on identifiability]{kadane1974role}.

\section{Log Restricted Likelihood and Prediction Formulas ~}\label{sec:predictions}

The log restricted likelihood for our model under the assumption of Gaussianity for $\Yvec$ can be written as \citep{cressie1993asymptotic},
\begin{equation*}\label{eq:reml}
	\mathcal{L}(\thetavec; \Zvec) = -\frac{N - pq}{2} \log(2 \pi) + \frac{1}{2} \log \abs{\Xvec' \Xvec} - \frac{1}{2} \log \abs{\Sigmavec_{Z}} - \frac{1}{2} \log \abs{\Xvec' \Sigmavec_{Z}^{-1} \Xvec} - \frac{1}{2} \Zvec' \Pivec \Zvec,
\end{equation*}
where $N = \sum_{i=1}^{p} n_i$, and
\( \Pivec = \Sigmavec_{Z}^{-1} - \Sigmavec_{Z}^{-1} \Xvec (\Xvec' \Sigmavec_{Z}^{-1} \Xvec)^{-1} \Xvec' \Sigmavec_{Z}^{-1}. \)

The estimate of $\betavec$, $\hat{\betavec} $, is given by
\begin{equation*}\label{eq:beta_eq}
	\hat{\betavec} = (\Xvec' \hat{\Sigmavec}_{Z}^{-1} \Xvec)^{-1} \Xvec' \hat{\Sigmavec}_{Z}^{-1} \Zvec,
\end{equation*}
where $\hat{\Sigmavec}_{Z}$ denotes $\Sigmavec_{Z}$ evaluated at $\thetavec = \hat \thetavec$.

Treating the plug-in REML estimates as known parameters, the joint distribution of the data $\Zvec$ and the process $\Yt_i(\cdot)$ evaluated at $\svec^{*}$ is
\begin{equation}\label{eq:new_y}
	\begin{pmatrix}
		\Zvec \\
		\Yt_i(\svec^{*})
	\end{pmatrix}
	\sim
	\Gau \left( \begin{pmatrix}
		\Xvec \\
		\xvec_i^{*'}
	\end{pmatrix} \betavec, \begin{pmatrix}
		\Sigmavec_{Z} & \sigmavec^{*}(\svec^{*}) \\
		\sigmavec^{*}(\svec^{*})' & C_{ii, G}(\svec^{*}, \svec^{*})
	\end{pmatrix} \right),
\end{equation}
where $\xvec_i^{*} = (\xvec(\svec^{*})' I(j=i): \quad j = 1,\dots,p)'$, and $\sigmavec^{*}(\svec^{*}) = (C_{1i, G}(\svec_{11}, \svec^{*}),$ $\dots,C_{1i, G}(\svec_{1 n_{1}}, \svec^{*}), $ $\dots, C_{pi, G}(\svec_{p1}, \svec^{*}),\dots,C_{pi, G}(\svec_{p n_{p}}, \svec^{*}))'$. From \eqref{eq:new_y}, Gaussian conditioning yields
\begin{align}\label{eq:pred}
	\begin{split}
		&\E(\Yt_i(\svec^{*}) \mid \Zvec) = \xvec_i^{*'} \betavec + \sigmavec^{*}(\svec^{*})' \Sigmavec_{Z}^{-1} (\Zvec - \Xvec \betavec), \\
		&\Var(\Yt_i(\svec^{*}) \mid \Zvec) = C_{ii, G}(\svec^{*}, \svec^{*}) - \sigmavec^{*}(\svec^{*})' \Sigmavec_{Z}^{-1} \sigmavec^{*}(\svec^{*}).
	\end{split}
\end{align}
From \eqref{eq:pred}, it is also straightforward to make a probabilistic prediction of an observation at $\svec^{*}$, say $Z_i^{*}$, since
$\E(Z_i^{*} \mid \Zvec) = \E(\Yt_i(\svec^{*}) \mid \Zvec)$, and
$\Var(Z_i^{*} \mid \Zvec) = \Var(\Yt_i(\svec^{*}) \mid \Zvec) + \tau^2_i$.

\section{Additional Tables and Figures ~}\label{sec:additional_results}

\begin{table}
	\centering
	\caption{Summary of the warping units. In each row, a warping unit is described. Parameters appearing in the functions composing the units are denoted using $\thetavec$. }
	\label{tbl:warping_units}
	\bgroup
	\def\arraystretch{1}
	\begin{tabular}{ |>{\centering\arraybackslash}m{3cm}|>{\centering\arraybackslash}m{6cm}|>{\centering\arraybackslash}m{4cm}|>{\centering\arraybackslash}m{3cm}| }
		\hline
		\textbf{Type of deformation function} & \textbf{Functional form} & \textbf{Usage} & \textbf{Visualization} \\
		\hline
		Axial warping & $\tilde \fvec(\svec) = \begin{pmatrix}\tilde f(s_1) \\ s_2 \end{pmatrix}$ or $\begin{pmatrix}s_1 \\ \tilde  f(s_2) \end{pmatrix}$, where $\tilde f(s) = \sum_{i = 1}^{r} w_i \phi_i(s)$, and where $\phi_1(s) = s$; $\phi_i(s) = \frac{1}{1 + \exp{-\theta_1(s-\theta_2)}}, i = 2, \dots, r$.  & warp space along one of the axes & \includegraphics[width=0.15\textwidth]{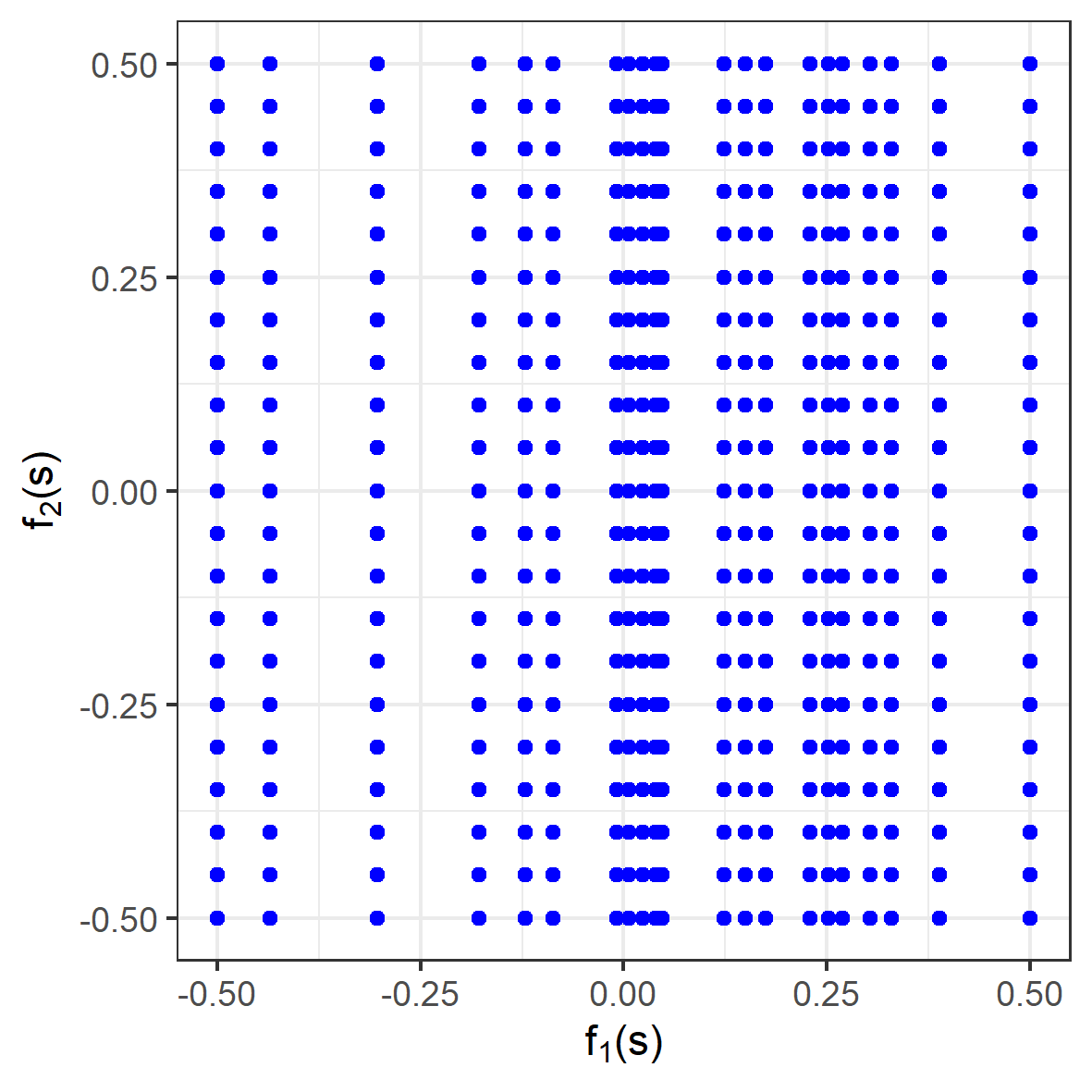} \\
		\hline
		Radial basis function & $\tilde \fvec(\svec) = \svec + w \exp{-\thetavec_2 \norm{\svec - \thetavec_1}^2} (\svec - \thetavec_1)$  & expand or contract space locally around the center of the basis function & \includegraphics[width=0.15\textwidth]{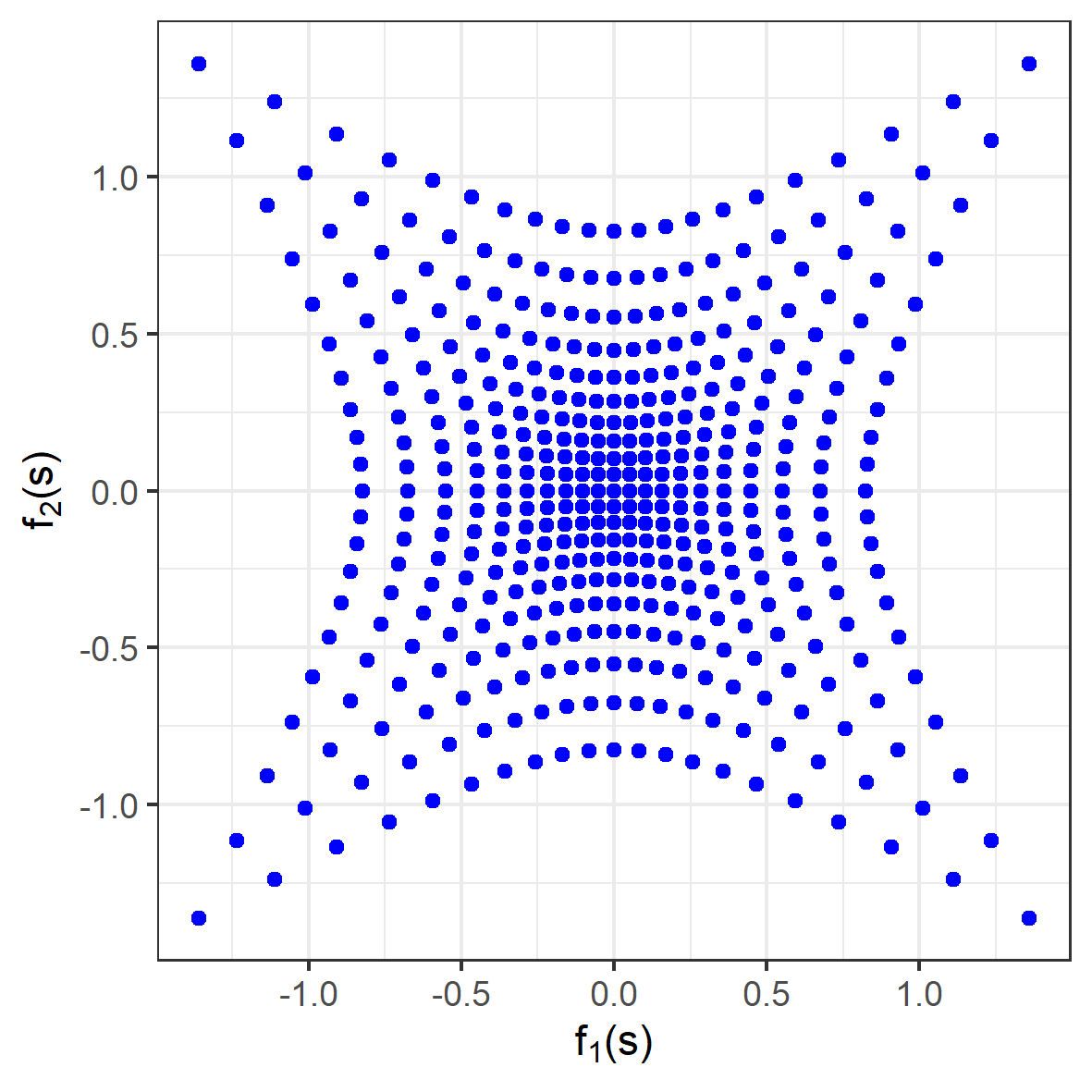} \\
		\hline
		M{\"o}bius transformation & $\tilde \fvec(\svec) = \begin{pmatrix} \Re(\phi(\svec)) \\ \Im(\phi(\svec)) \end{pmatrix}$, where $\phi(\svec) = \frac{\theta_1 z(\svec) + \theta_2}{\theta_3 z(\svec) + \theta_4}$;  $z(\svec) = s_1 + i s_2 ; \theta_1, \theta_2, \theta_3, \theta_4 \in \mathbb{C}$  & move points around fixed points (usually in circular paths) & \includegraphics[width=0.15\textwidth]{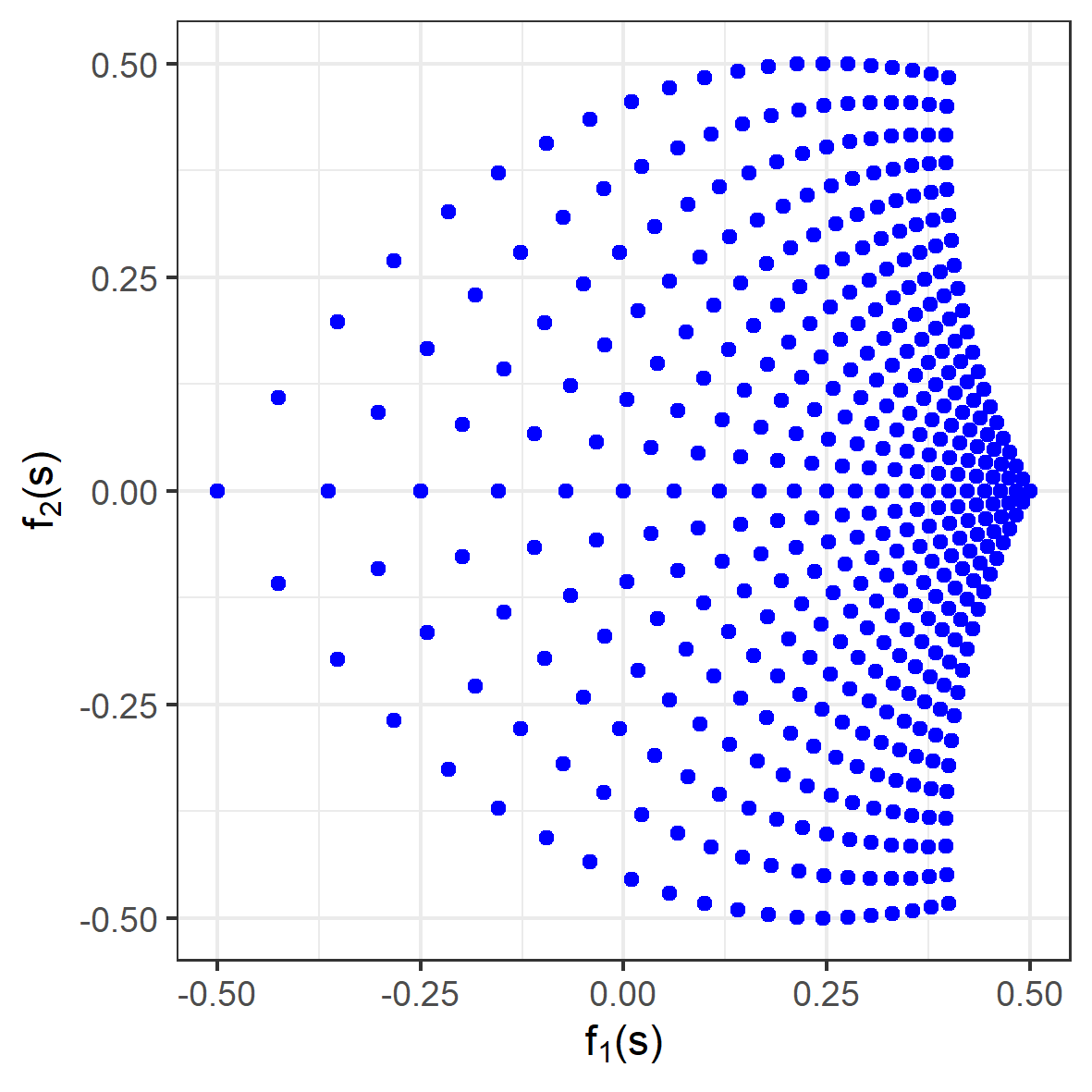} \\
		\hline
		Affine transformation & $\tilde \gvec(\svec) = \Avec \svec + \dvec$  & align processes with respect to the first process (using shifts and rotations) & \includegraphics[width=0.15\textwidth]{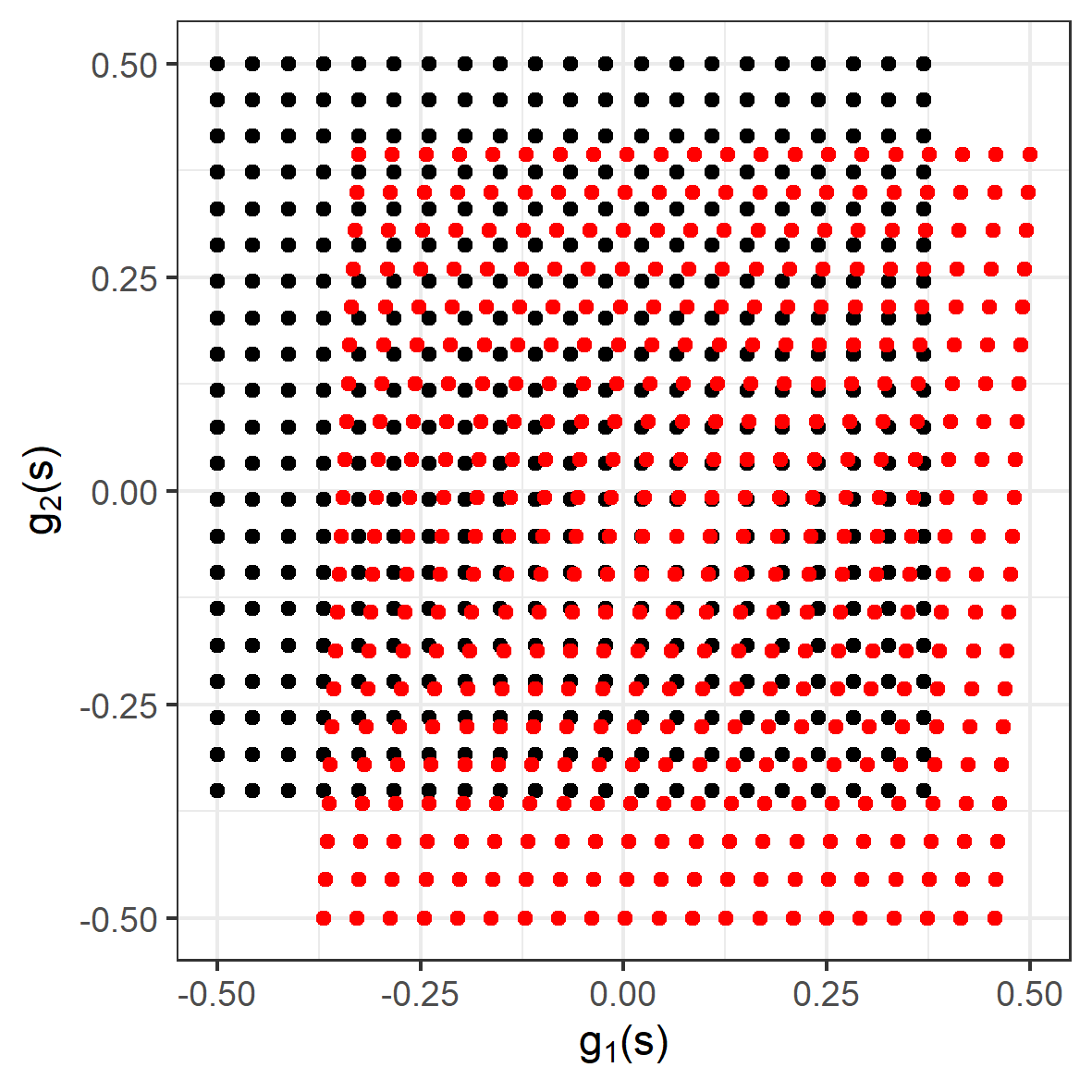} \\
		\hline
	\end{tabular}
	\egroup
\end{table}

\begin{figure}
	\centering
	
	\includegraphics[width=0.8\textwidth]{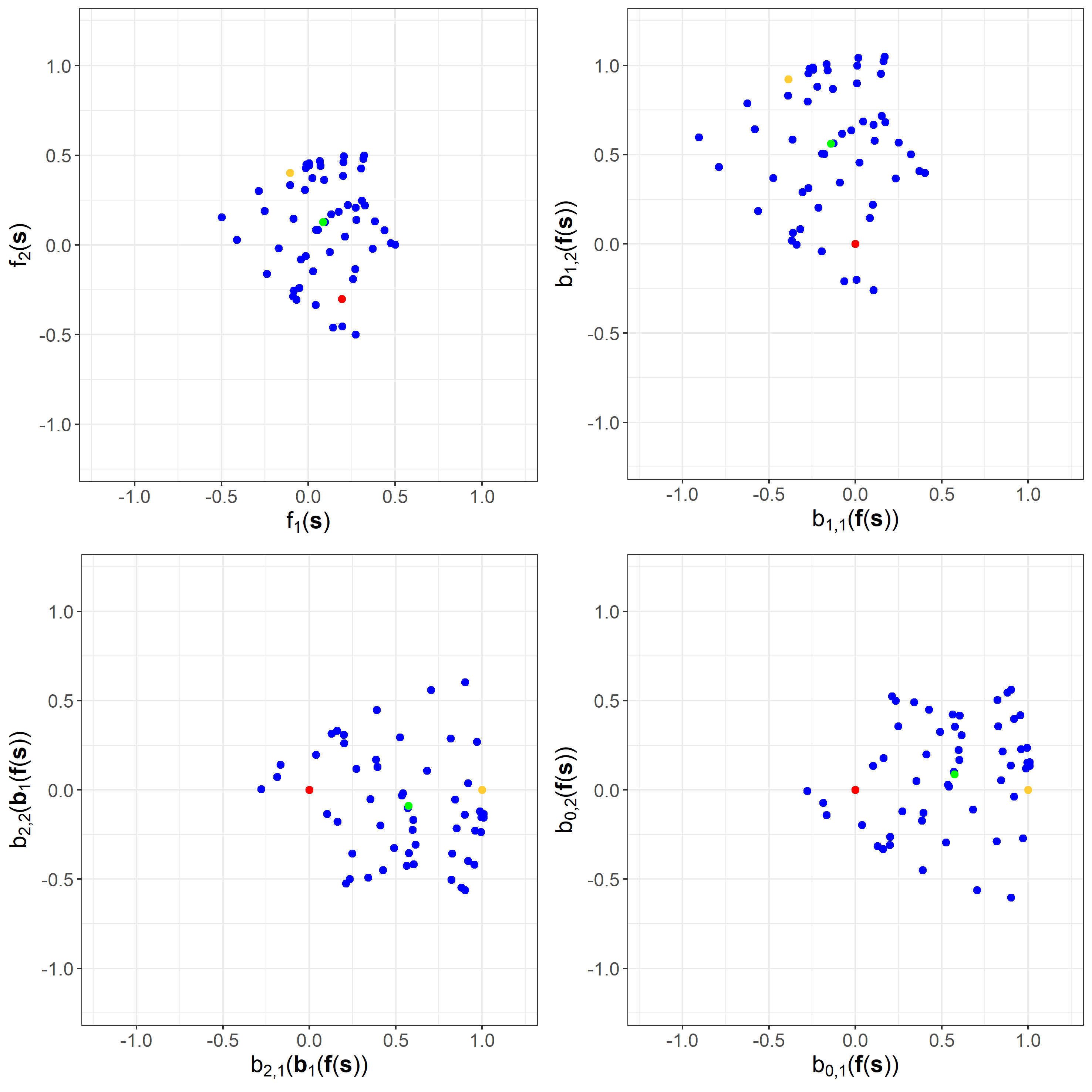}
	
	\caption{Illustration of the homogenizing function $\bvec_0(\cdot)$. Top left: A set of locations on the warped domain $D$, with the red, yellow, and green dots denoting $\fvec(\svec_k), \fvec(\svec_l)$, and $\fvec(\svec_m)$, respectively. Top right: The locations on the scaled and shifted domain $D_1$. Bottom left: The locations on the scaled, shifted, and rotated domain $D_2$. Bottom right: \Copy{rev3iip14}{The locations on the scaled, shifted, rotated, and reflected, domain $D_3$. This is the fixed frame of reference defined in Section \ref{sec:frame_of_ref}.} }
	\label{fig:warping_viz}
\end{figure}

\begin{algorithm}
	\nl Find the REML estimate $\hat{\thetavec}$ by maximizing $\mathcal{L}$ in \eqref{eq:reml}. \\
	\nl Find the REML estimate $\hat{\betavec}$ using \eqref{eq:beta_eq}. \\
	\nl Find the Cholesky factor $\Lvec$ of $\hat{\Sigmavec}_{Z}$, that is, find $\Lvec$ such that $\Lvec \Lvec' = \hat{\Sigmavec}_{Z}$. \\
	\nl Decorrelate the data: $\Zvec_0 = \Lvec^{-1} (\Zvec - \Xvec \hat{\betavec})$. \\
	\nl Generate an uncorrelated bootstrap sample $\Zvec_{0,b}$ by sampling $\Zvec_0$ with replacement. \\
	\nl Create the correlated bootstrap sample $\Zvec_{b} = \Lvec \Zvec_{0,b} + \Xvec \hat{\betavec}$. \\
	\nl Find the bootstrap estimate $\hat{\thetavec}_b$ and $\hat{\betavec}_b$ from $\Zvec_{b}$. \\
	\nl Repeat from step 5 for $B$ times to create a bootstrap sample of size $B$. (For most problems, $B \approx 1000$ should suffice.)
	
	\caption{{\bf Parameter bootstrapping for uncertainty quantification} \label{alg:bootstrap}}
	
\end{algorithm}

\begin{figure}
	\centering
	\includegraphics[width=1\textwidth]{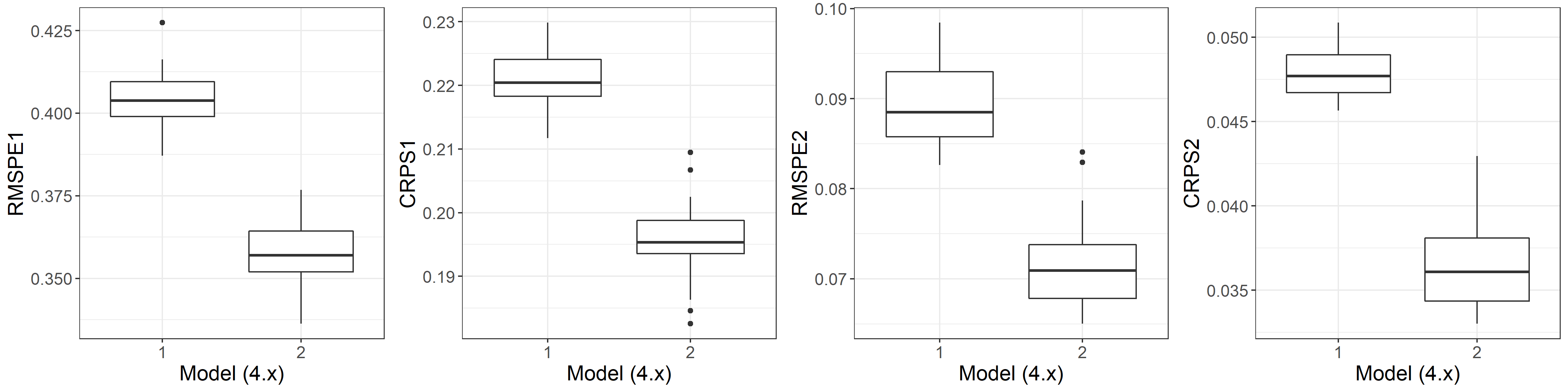}
	\caption{
		\footnotesize{Boxplots of RMSPE and CRPS for both models across 30 simulations in Section \ref{sec:simstudies_sym}. The left two panels correspond to $\tilde Y_1(\cdot)$, and the right two panels to $\tilde Y_2(\cdot)$.}
	}
	\label{fig:sim_symm_boxplot}
\end{figure}

\begin{table}
	\centering
	\caption{True values, estimates, and 95\% bootstrap confidence intervals (CI), of the model parameters for the simulation study in Section \ref{sec:simstudies_sym}, where the measurement locations are randomly sampled from $G$. }
	\label{tbl:est_par}
	\bgroup
	\def\arraystretch{1}
	\begin{tabular}{ |c|c|c|c| }
		\hline
		Parameters & True values & REML estimates & 95\% bootstrap CI \\
		\hline
		$\nu_{11}$ & 0.500 &	0.528 & (0.235, 0.634) \\
		\hline
		$\nu_{22}$ & 1.500 &	1.226 & (0.929, 1.664) \\
		\hline
		$\sigma_1$ & 1.000 &	1.040 & (0.775, 1.329) \\
		\hline
		$\sigma_2$ & 0.900 &	0.932 & (0.763, 1.388) \\
		\hline
		$\rho_{12}$ & 0.450 &	0.392 & (0.321, 0.537) \\
		\hline
		1/$\tilde{a}$ & 0.329 &	0.412 & (0.274, 0.556) \\
		\hline
		$\tau_1$ & 0.200 &	0.252 & (0.178, 0.265) \\
		\hline
		$\tau_2$ & 0.100 &	0.097 & (0.092, 0.108) \\
		\hline
		$\beta_{11}$ & 0.000 &	-0.232 & (-0.947, 0.415) \\
		\hline
		$\beta_{21}$ & 0.000 &	0.047 & (-0.762, 0.741) \\
		\hline
	\end{tabular}
	\egroup
\end{table}

\begin{table}
	\centering
	\caption{
		Five-fold cross-validation results, AIC, and the time required to fit, for the ocean temperatures at depths 0.5 meters and 318.1 meters for the study in Section \ref{sec:applications_ocean}, where data were missing at random.	}
	\label{tbl:crossvalid_ocean_temp}
	\bgroup
	\def\arraystretch{1}
	\begin{tabular}{ |c|c|c|c|c|c|c| }
		\hline
		& \multicolumn{2}{c|}{$T_{0.5}$} & \multicolumn{2}{c|}{$T_{318.1}$} & & \\
		\cline{2-5}
		& RMSPE & CRPS & RMSPE & CRPS & AIC & Time (s) \\
		\hline
		Model 4.2.1 & 0.0661 & 0.0306 & 0.0265 & 0.0129 & -4791.1 & 1338.8 \\
		\hline
		Model 4.2.2 & 0.0584 & 0.0188 & 0.0280 & 0.0136 & -6246.4 & 2545.8 \\
		\hline
		Model 4.2.3 & 0.0666 & 0.0198 & 0.0241 & 0.0123 & -6403.4 & 4455.2 \\
		\hline
	\end{tabular}
	\egroup
\end{table}

\begin{figure}
	\centering
	\includegraphics[width=0.6\textwidth]{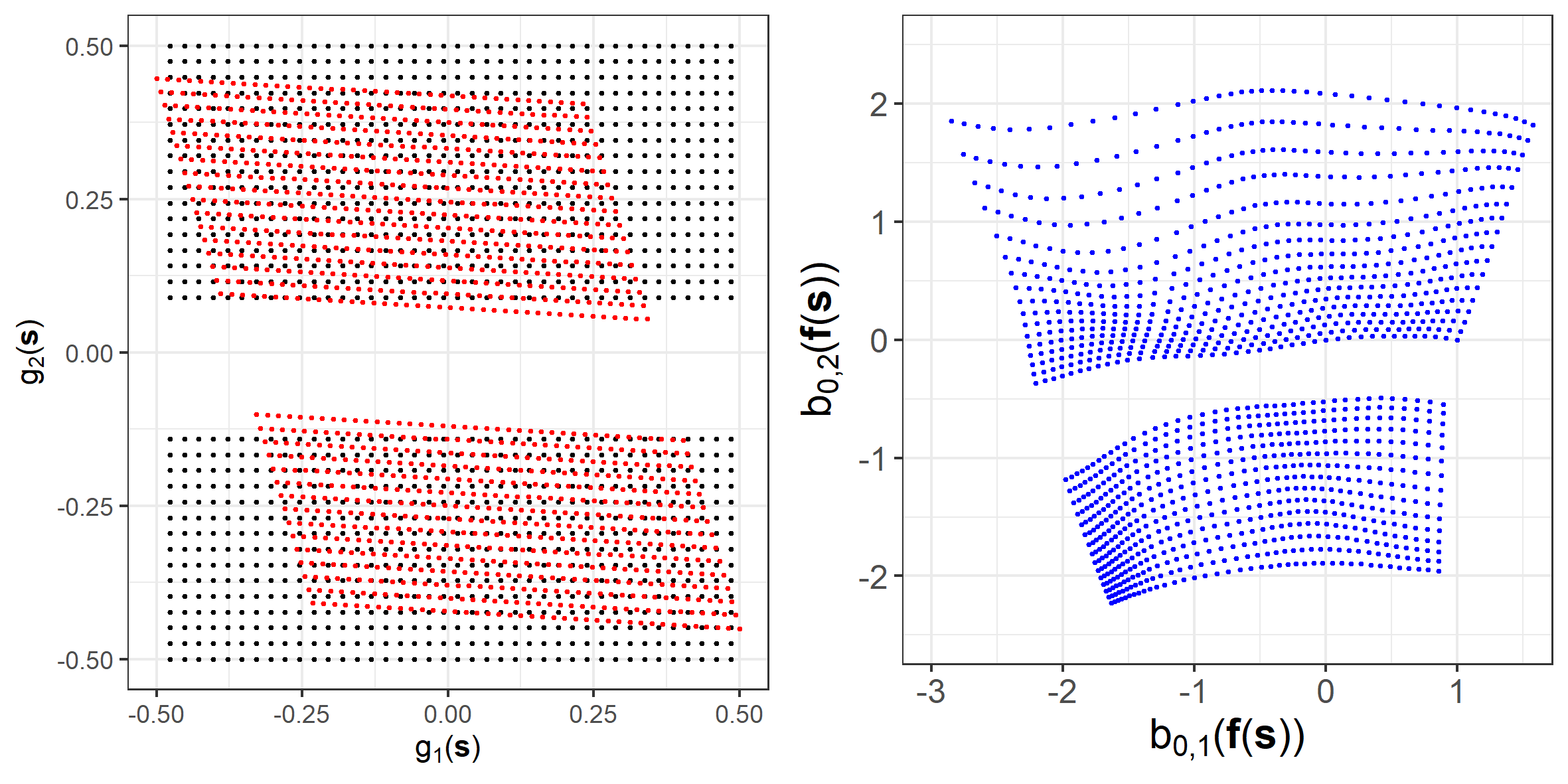}
	\caption{
		\footnotesize{Measurement locations under the estimated aligning function (left panel) and the homogenized warping function (right panel) for the ocean-temperatures data set in Section \ref{sec:applications_ocean}, where data are missing in a block.}
	}
	\label{fig:ocean_warping}
\end{figure}

\begin{figure}
	\centering
	\includegraphics[width=0.6\textwidth]{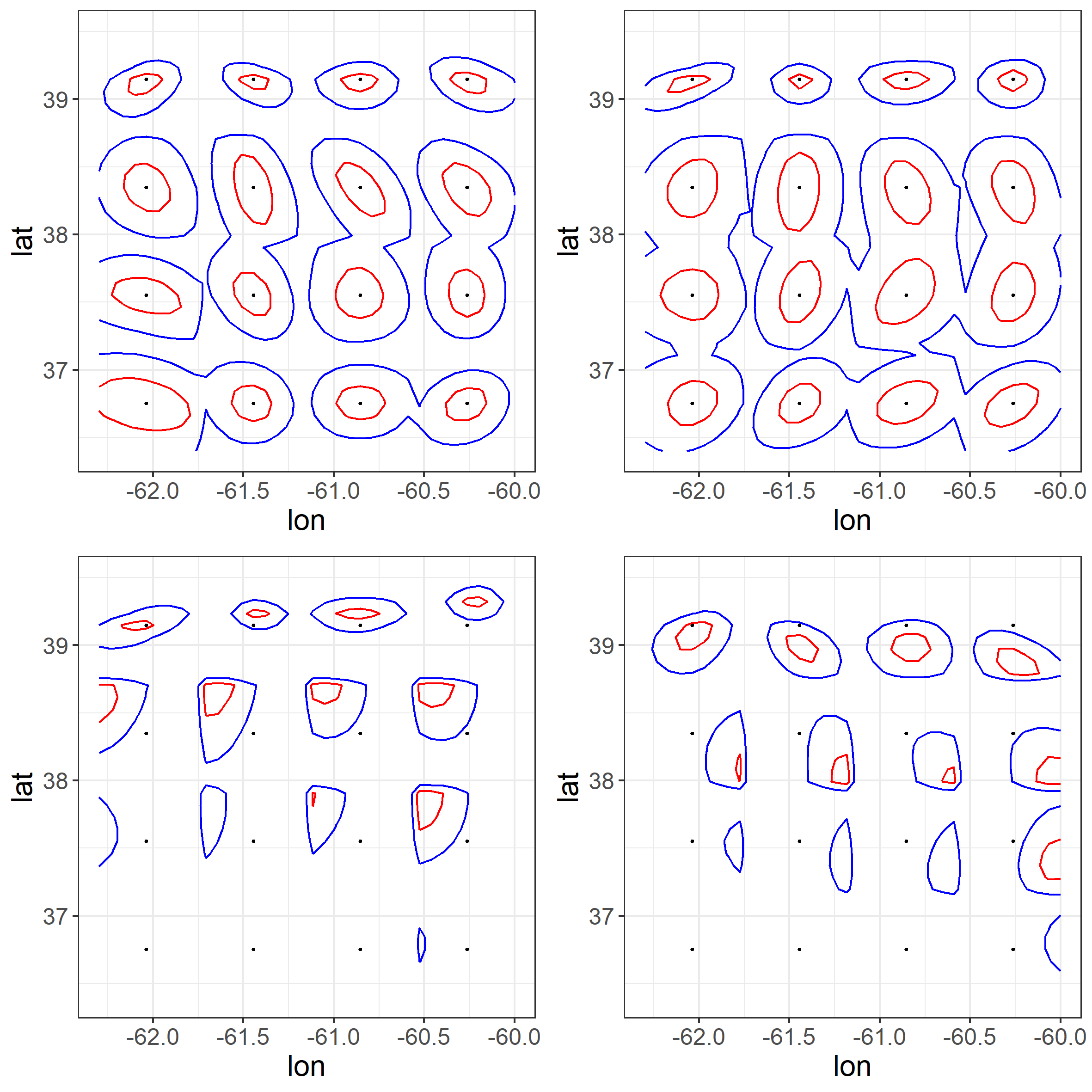}
	\caption{
		\footnotesize{Contours illustrating the estimated covariance functions evaluated at a set of locations (black dots) for the data set in Section \ref{sec:applications_ocean}. Blue and red contours denote covariances equal to 0.4 and 0.8 times of $\hat \sigma_{ij}^2$, where $\hat \sigma_{ij}^2 = \hat \sigma_{i}^2$, for $i = j$, and $\hat \sigma_{ij}^2 = \hat \rho_{ij} \hat \sigma_{i} \hat \sigma_{j}$, for $i \neq j$, respectively. Top row: Marginal covariances of the first process, $C_{11, G}(\cdot, \cdot)$ (left) and second process, $C_{22, G}(\cdot, \cdot)$ (right). Bottom row: Cross-covariances of the first process with the second process, $C_{12, G}(\cdot, \cdot)$ (left) and of the second process with the first process, $C_{21, G}(\cdot, \cdot)$ (right). Note the asymmetry. }
	}
	\label{fig:ocean_heatmap}
\end{figure}

\newpage

\section{Additional Data Illustrations ~}\label{sec:additional_illustrations}

\subsection{Simulated Symmetric Nonstationary Data With A Missing Block ~}\label{sec:simstudies_sym_block}

\begin{figure}
	\centering
	
	\includegraphics[width=0.8\textwidth]{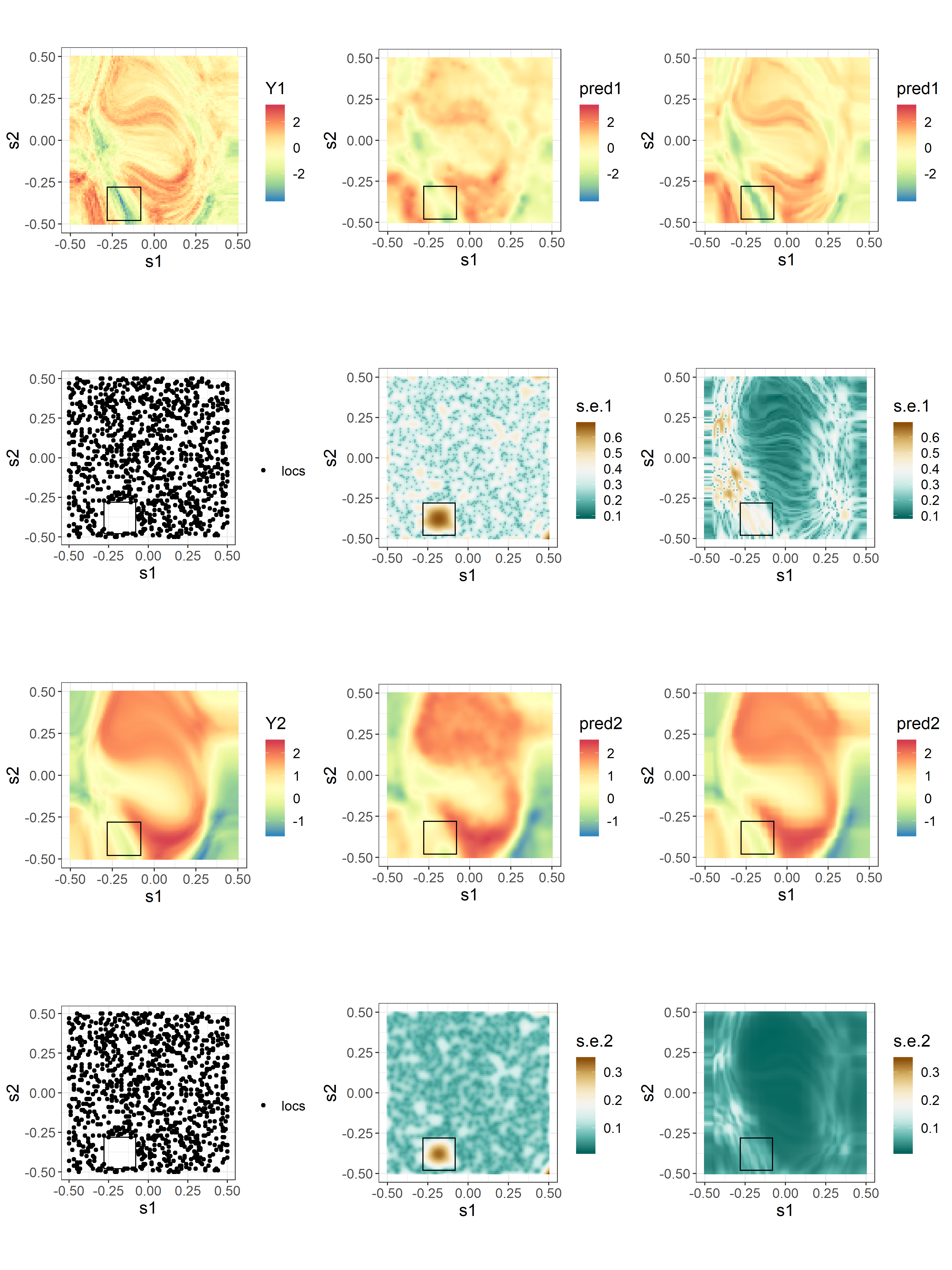}
	
	\caption{Same as Figure~\ref{fig:sim_study}, but where the measurement locations are randomly sampled from $G \backslash G_0$, and where the hold-out region $G_0$ is enclosed by the black square.}
	\label{fig:sim_study_gap}
\end{figure}

Following the simulation study in Section \ref{sec:simstudies_sym}, we considered the case where the symmetric nonstationary data are missing in a block. This situation occurs often when observing environmental variables (for example, clouds could prevent a remote sensing instrument from collecting data over a large region). \Copy{rev2p2.3}{As in the study in Section \ref{sec:simstudies_sym}, we sampled 1000 measurement locations at random 30 times, but this time on $G \backslash G_0$, where  $G_0 \equiv [-0.28, -0.08] \times [-0.48, -0.28]$ (i.e., the block of data was omitted for both processes).} Model 4.1.1 and Model 4.1.2 were then fitted to the data. Figure \ref{fig:sim_study_gap} shows the true simulated fields, the predictions, and the prediction standard errors from the two models. From Figure \ref{fig:sim_study_gap}, we see that the predictions from the DCSM recover the salient features in the true fields despite the relatively large gap. The DCSM also produces relatively lower prediction standard errors in the unobserved region than the stationary parsimonious Mat{\'e}rn model.
Table \ref{tbl:sim_gap} shows the RMSPE and CRPS from the two models when predicting the latent process at the grid locations in $G_0$ and, again, it illustrates the improvement in RMSPE and CRPS that can be achieved when accounting for complex nonstationary properties of the process, even when the data have large gaps.

\begin{table}
	\centering
	\caption{Average hold-out-validation results, AIC, and the time required to fit, for the simulation study in Section \ref{sec:simstudies_sym_block}, where the measurement locations are randomly sampled 30 times from $G \backslash G_0$.}
	\label{tbl:sim_gap}
	\bgroup
	\def\arraystretch{1}
	\begin{tabular}{ |c|c|c|c|c|c|c| }
		\hline
		& \multicolumn{2}{c|}{$\tilde Y_1(\cdot)$} & \multicolumn{2}{c|}{$\tilde Y_2(\cdot)$} & & \\
		\cline{2-5}
		& RMSPE & CRPS & RMSPE & CRPS & AIC & Time (s) \\
		\hline
		Model 4.1.1 & 1.137 &	0.675  & 0.453 & 0.274 & 833.2 & 816.9 \\
		\hline
		Model 4.1.2 & 0.713 &	0.424  & 0.261 & 0.167 & 310.4 & 1533.0  \\
		\hline
	\end{tabular}
	\egroup
\end{table}

\subsection{Simulated Asymmetric Nonstationary Data ~}\label{sec:simstudies_asym}

We now demonstrate inference for a bivariate DCSM from simulated asymmetric nonstationary data. We simulated bivariate data from a bivariate Gaussian DCSM $\tilde \Yvec(\cdot)$ with constant mean (i.e., $q = 1$ and $\xvec(\cdot) = x(\cdot) = 1$) in \eqref{eq:y_model}, and cross-covariance matrix function as in \eqref{eq:cov_model_asym}. The data were simulated on the 101 $\times$ 101 grid defined on the geographic domain $G$. Now the \emph{shared warping function} $\fvec(\cdot)$ is a composition of axial warping units, a single resolution radial basis function unit, and a M{\"o}bius transformation unit, while the \emph{aligning function} $\gvec_2(\cdot)$ is an affine transformation. (Recall from Section \ref{sec:Model_multi_main} that we fix $\gvec_1(\cdot)$ to be the identity map.) On the warped domain, we use a stationary bivariate parsimonious Mat{\'e}rn model. As in Section \ref{sec:simstudies_sym}, we randomly sampled without replacement 1000 locations from the grid and used them as measurement locations.

We compared the predictive performance of the following five models on the $101 \times 101$ grid on $G$.
\begin{itemize}
	\item Model S4.2.1: A bivariate, stationary, symmetric, parsimonious Mat{\'e}rn model.
	\item Model S4.2.2: A bivariate, marginally stationary, asymmetric model with $\fvec(\cdot)$ the identity map and the aligning function $\gvec_2(\cdot)$ an affine transformation (as described in Proposition \ref{propo:marginal_stat_model}),
	with Model S4.2.1 on the warped domain. 
	\item Model S4.2.3: A univariate DCSM for each of the processes, with the warping function $\fvec(\cdot)$ a composition of axial warping units, a single-resolution radial basis function, and a M{\"o}bius transformation unit, with the Mat{\'e}rn covariance model on the warped domain.
	\item Model S4.2.4: A bivariate symmetric DCSM, with the warping function as in Model S4.2.3, with Model S4.2.1 on the warped domain.
	\item Model S4.2.5: A bivariate asymmetric DCSM, with the aligning function $\gvec_2(\cdot)$ as in Model S4.2.2, the warping function as in Model S4.2.3, and Model S4.2.1 on the warped domain. This is the model from which the data were simulated.
\end{itemize}

\begin{figure}
	\centering
	
	\includegraphics[width=1\textwidth]{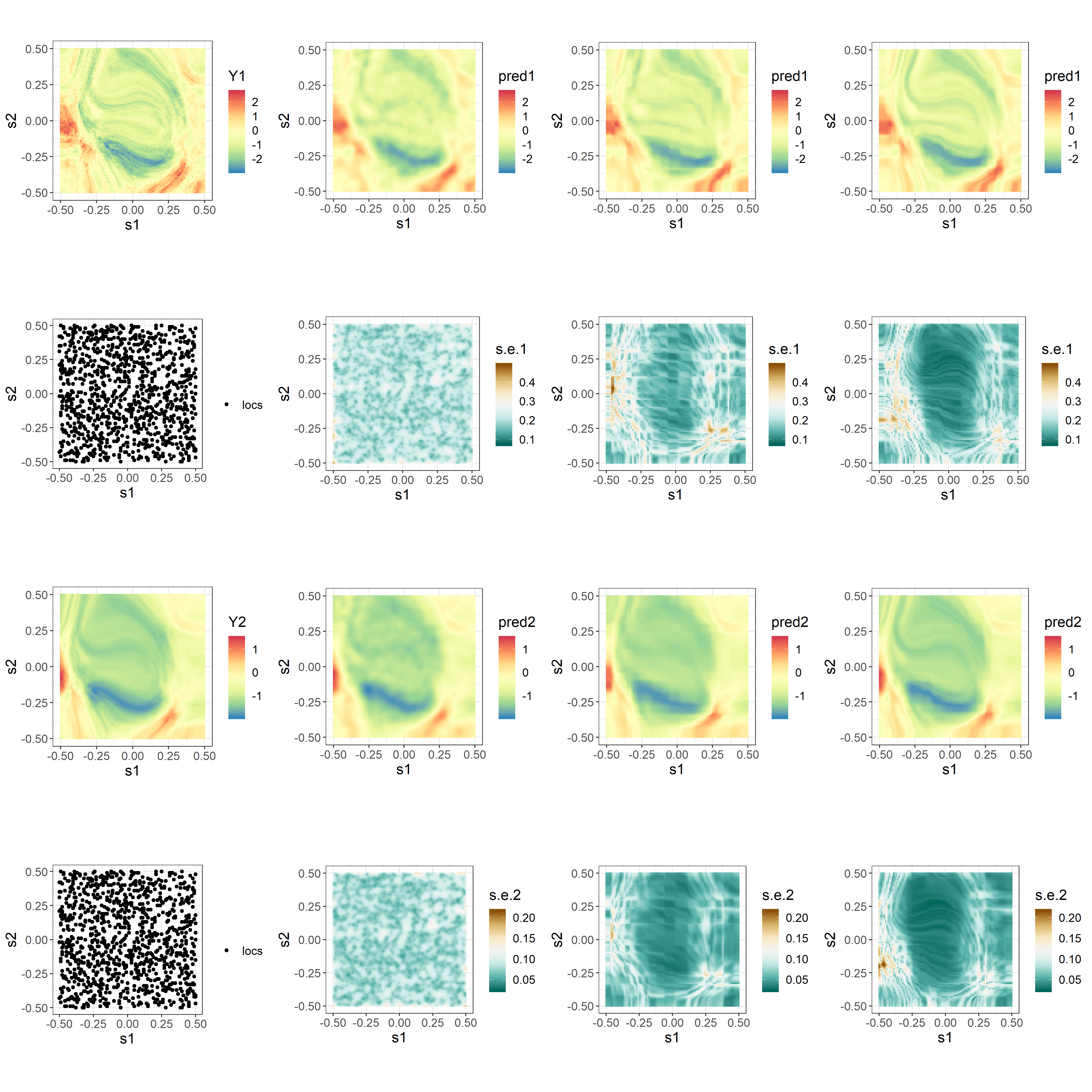}
	
	\caption{
		Comparison of predictions and prediction standard errors when using an asymmetric, stationary, parsimonious Mat{\'e}rn model (Model S4.2.2); a symmetric DCSM (Model S4.2.4); and an asymmetric DCSM (Model S4.2.5) in the study of Section \ref{sec:simstudies_asym}, where measurement locations were randomly sampled without replacement from $G$.
		First row: The process $\tilde Y_1(\cdot)$ (first panel); predictions obtained using Model S4.2.2 (second panel), Model S4.2.4 (third panel) and Model S4.2.5 (fourth panel). Second row: Locations of the measurement of $\tilde Y_1(\cdot)$ (first panel); prediction standard errors obtained when using Model S4.2.2 (second panel), Model S4.2.4 (third panel) and Model S4.2.5 (fourth panel). Third and fourth rows: Analogous to the first and second rows, respectively, for the process $\tilde Y_2(\cdot)$.
	}
	\label{fig:sim_study_asym}
\end{figure}

Figure \ref{fig:sim_study_asym} shows the true simulated fields and the predicted fields from Model S4.2.2, Model S4.2.4, and Model S4.2.5. From Figure \ref{fig:sim_study_asym}, we can see that Model S4.2.2 smooths out certain features (similar to the symmetric case), while Model S4.2.5 is able to reproduce sharper features than Model S4.2.3, illustrating that both nonstationarity and asymmetry could be important when modeling multivariate spatial processes. Figure \ref{fig:sim_study_asym} also shows the prediction standard errors for Model S4.2.2, Model S4.2.4, and Model S4.2.5. As in Section \ref{sec:simstudies_sym}, we see that while there is no pattern in the prediction-standard-error map for Model S4.2.2, the DCSMs produce prediction standard errors that are reflective of the processes' local anisotropies and scales.

\begin{figure}[t!]
	\centering
	\includegraphics[width=0.6\textwidth]{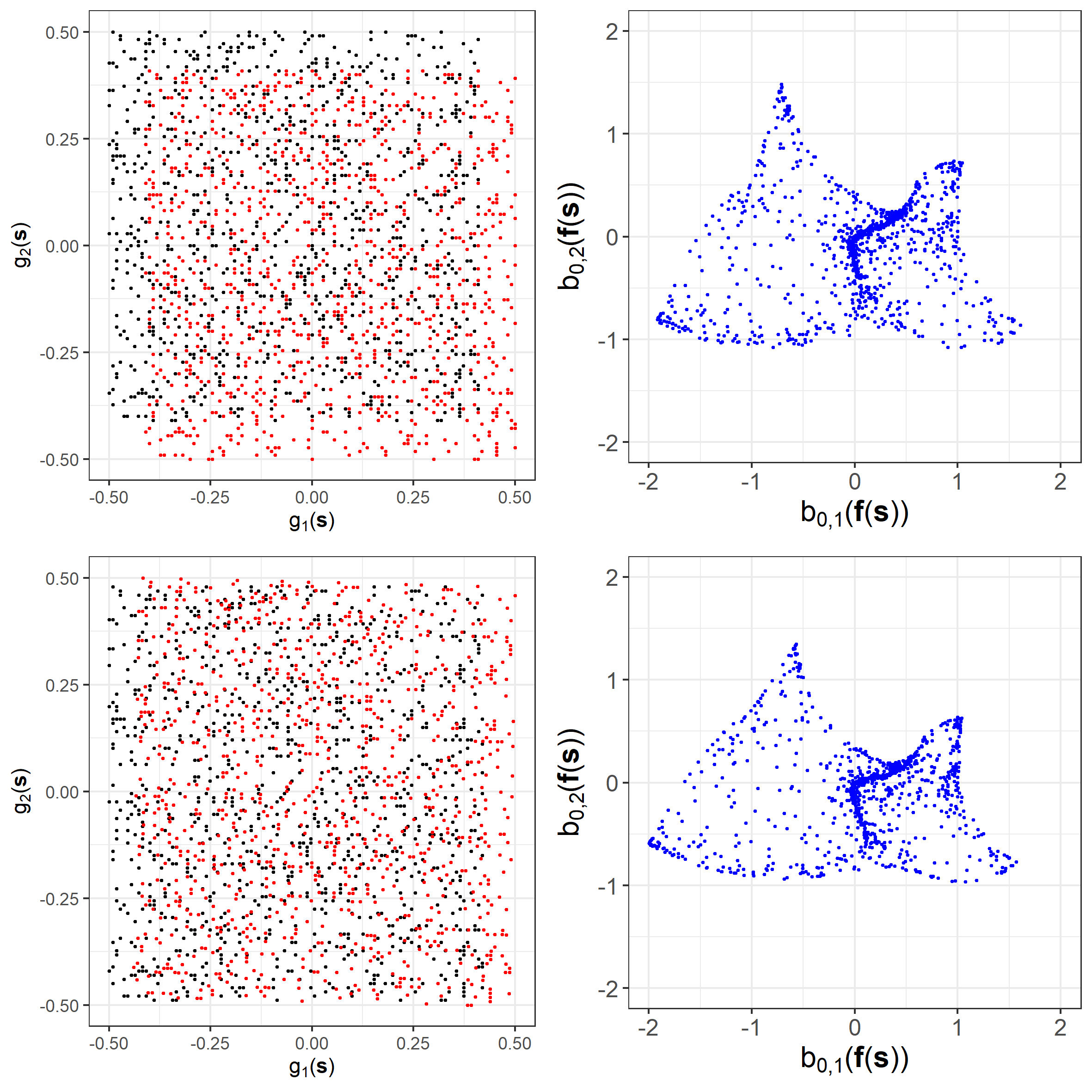}
	\caption{
		\footnotesize{Measurement locations under the aligning functions and the homogenized warping functions for the data set in Section \ref{sec:simstudies_asym}. Top row: Measurement locations under the true aligning function (left panel) and the true warping function (right panel). Bottom row: Measurement locations under the estimated aligning function (left panel) and the estimated warping function (right panel) using Model S4.2.5.}
	}
	\label{fig:warped_loc_asym}
\end{figure}

We can also compare, respectively, the estimated aligning function and the estimated warping function in Model S4.2.5 to the true aligning function and the true warping function in Model S4.2.5. Figure \ref{fig:warped_loc_asym} shows the measurement locations under the true aligning and warping functions, and the measurement locations under the estimated aligning and warping functions. \Copy{rev1p2}{We see that the estimated aligning function generates a shift towards the east direction, which is similar to the true aligning function, which generates a shift towards the southeast direction. One can interpret that the second process needs to be shifted eastwards to align with the first process. The estimated warping function also retains important features of the true warping function, such as the contraction in the middle part of the domain. This can be interpreted that on the original domain $G$, the scale parameter in the middle region is smaller than the scale parameter in the boundary region.}

\begin{figure}[t!]
	\centering
	\includegraphics[width=0.6\textwidth]{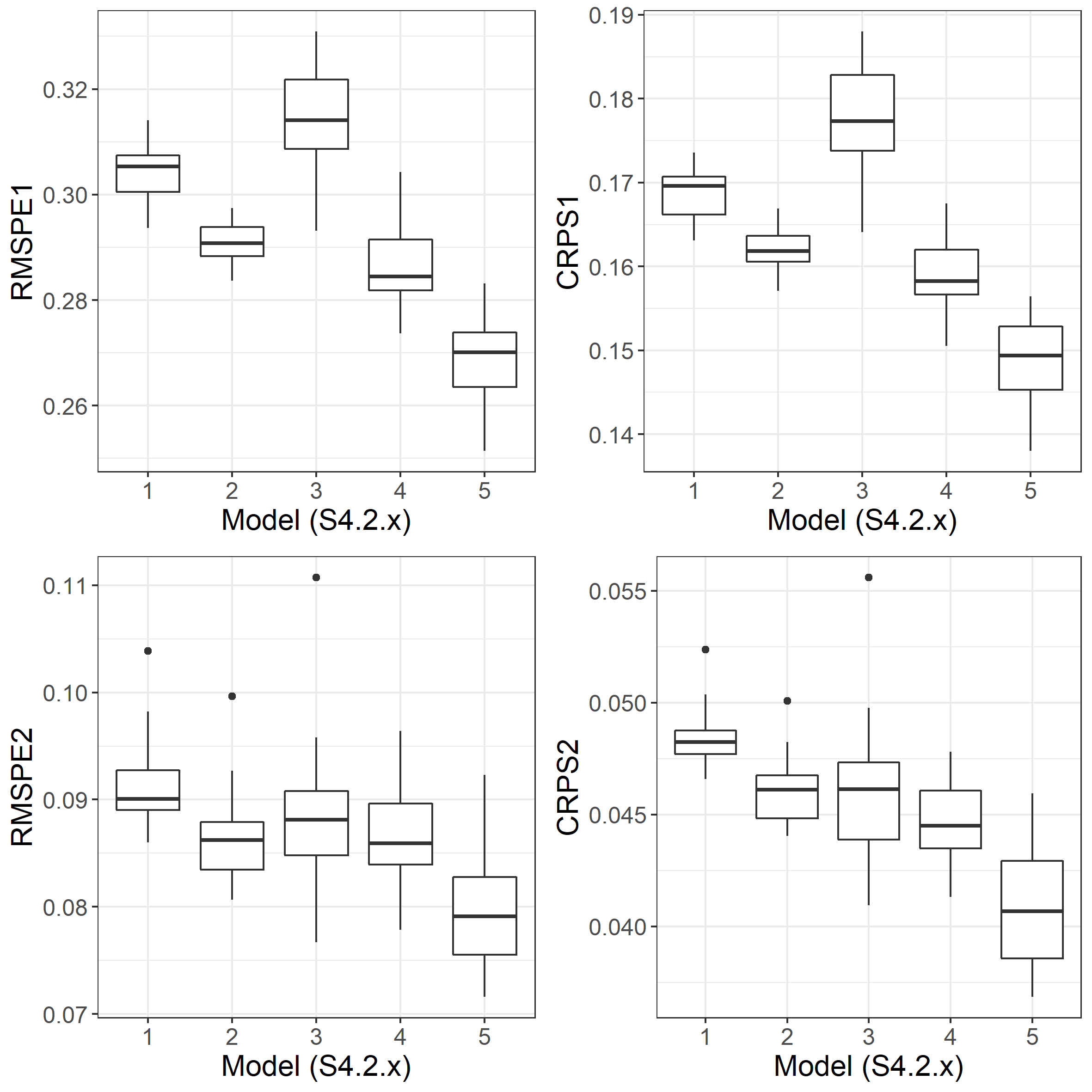}
	\caption{
		\footnotesize{Boxplots of RMSPE and CRPS for the five models across 30 simulations in Section \ref{sec:simstudies_asym}. The top row corresponds to $\tilde Y_1(\cdot)$, and the bottom row to $\tilde Y_2(\cdot)$.}
	}
	\label{fig:sim_asymm_boxplot}
\end{figure}

In a similar manner to Section \ref{sec:simstudies_sym}, we repeated the procedure of randomly sampling 1000 locations 30 times from $G$ and computing predictive diagnostics. Table \ref{tbl:sim_asym} displays the results from the cross-validation study, and Figure \ref{fig:sim_asymm_boxplot} shows the boxplots of the RMSPE and CRPS for the models across the 30 simulations. Model S4.2.5, which considers both nonstationarity and asymmetry, produces the best predictions in terms of lowest RMSPE and CRPS, while the symmetric nonstationary model (Model S4.2.4) as well as the asymmetric stationary model (Model S4.2.2) are seen to yield slight improvements over the conventional symmetric, stationary model (Model S4.2.1). Surprisingly, even when accounting for nonstationarity in each process, the decoupled univariate model (Model S4.2.3) yields the worst predictions, showing the importance of a multivariate model.

\begin{table}
	\centering
	\caption{Average hold-out-validation results, AIC, and the time required to fit, for the simulation study in Section \ref{sec:simstudies_asym}, where the measurement locations are randomly sampled 30 times from $G$.}
	\label{tbl:sim_asym}
	\bgroup
	\def\arraystretch{1}
	\begin{tabular}{ |c|c|c|c|c|c|c| }
		\hline
		& \multicolumn{2}{c|}{$\tilde{Y}_1(\cdot)$} & \multicolumn{2}{c|}{$\tilde{Y}_2(\cdot)$} & & \\
		\cline{2-5}
		& RMSPE & CRPS  & RMSPE & CRPS & AIC & Time (s) \\
		\hline
		Model S4.2.1 & 0.304 &	0.169  & 0.091 & 0.048 & 386.3 & 815.3 \\
		\hline
		Model S4.2.2 & 0.291 &	0.162  & 0.086 & 0.046 & 195.4 & 1188.6 \\
		\hline
		Model S4.2.3 & 0.314 &	0.177  & 0.088 & 0.046 & 44.2 & 2025.7 \\
		\hline
		Model S4.2.4 & 0.287 &	0.159  & 0.087 & 0.045 & -3.8 & 1530.5 \\
		\hline
		Model S4.2.5 & 0.269 &	0.149  & 0.080 & 0.041 & -242.5 & 2654.1  \\
		\hline
	\end{tabular}
	\egroup
\end{table}

\subsection{Simulated Data with Misspecified Warping Functions ~}\label{sec:misspecified}

In this section, we consider cases where the deformations are misspecified. Specifically, we present two cases: First, we consider data simulated from a bivariate, symmetric, stationary covariance function. Second, we consider data simulated from a bivariate, asymmetric, nonstationary covariance function with a misspecified warping function.

\subsubsection{Simulated Symmetric Stationary Data}\label{sec:misspecified1}

In this section, we demonstrate the use of a bivariate DCSM with simulated symmetric stationary data. We simulated bivariate data from a bivariate, stationary, symmetric, parsimonious Mat{\'e}rn model. The data were simulated on a 101 $\times$ 101 grid defined on the geographic domain $G$. We randomly sampled without replacement 1000 locations from the grid and used them as measurement locations.

\begin{figure}[t!]
	\centering
	\includegraphics[width=0.6\textwidth]{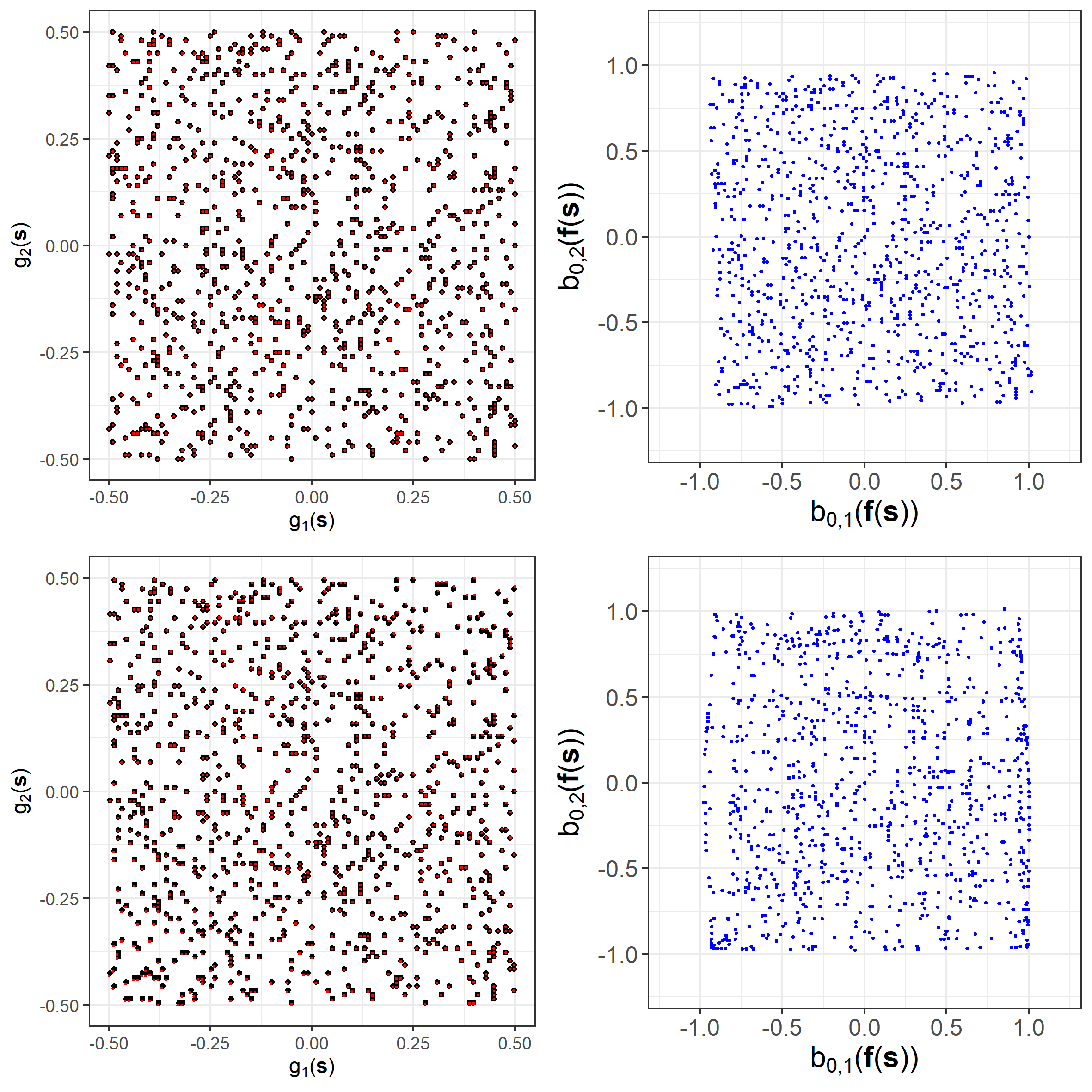}
	\caption{
		\footnotesize{Measurement locations under the aligning functions and the homogenized warping functions for the data set in Section \ref{sec:misspecified1}. Top row: Measurement locations under the true aligning function (left panel) and the true warping function (right panel). Bottom row: Measurement locations under the estimated aligning function (left panel) and the estimated warping function (right panel).}
	}
	\label{fig:warped_loc_sym_stat}
\end{figure}

We first examined the ability of the DCSM to retrieve the true warping function (in this case, the identity function). We used the same architecture in the DCSM as in Model S4.2.5 in Section \ref{sec:simstudies_asym}. Figure \ref{fig:warped_loc_sym_stat} shows the measurement locations under the true aligning and warping function, and the measurement locations under the estimated aligning and warping function. We observe that both the aligning function and the warping function have been correctly estimated to be approximately the identity functions.

We also compared the predictive performance of the DCSM with the true model.

\begin{itemize}
	\item Model S4.3.1.1: The bivariate, stationary, symmetric, parsimonious Mat{\'e}rn model. This is the model from which the data were simulated.
	\item Model S4.3.1.2: The same bivariate DCSM as Model S4.2.5 in Section \ref{sec:simstudies_asym}.
\end{itemize}

Table \ref{tbl:sim_misspecified1} shows the cross-validation results of the simulation study. As expected, Model S4.3.1.2 produces worse RMSPE and CRPS than Model S4.3.1.1, but it does not perform much worse. Given that we use very complex deformations in Model S4.3.1.2, this shows that the DCSM is quite robust to overfitting.

\begin{table}
	\centering
	\caption{Hold-out-validation results, AIC, and the time required to fit, for the simulation study in Section \ref{sec:misspecified1}, where the measurement locations are randomly sampled from $G$.}
	\label{tbl:sim_misspecified1}
	\bgroup
	\def\arraystretch{1}
	\begin{tabular}{ |c|c|c|c|c|c|c| }
		\hline
		& \multicolumn{2}{c|}{$\tilde{Y}_1(\cdot)$} & \multicolumn{2}{c|}{$\tilde{Y}_2(\cdot)$} & & \\
		\cline{2-5}
		& RMSPE & CRPS  & RMSPE & CRPS & AIC & Time (s) \\
		\hline
		Model S4.3.1.1 & 0.334 & 0.188  & 0.097 & 0.054 & 304.1 & 840.6 \\
		\hline
		Model S4.3.1.2 & 0.346 & 0.195  & 0.108 & 0.060 & 437.1 & 2714.9 \\ 
		\hline
	\end{tabular}
	\egroup
\end{table}

\subsubsection{Simuated Asymmetric Nonstationary Data with a Misspecified Warping Function}\label{sec:misspecified2}

In this section, we demonstrate the use of a DCSM with simulated asymmetric nonstationary data with a misspecified warping function. We simulated bivariate data from a bivariate Gaussian DCSM in a manner similar to Section \ref{sec:simstudies_asym}, but now we use a different warping function, $\fvec(\svec) = \ovec + (\svec - \ovec) \norm{\svec - \ovec}$ \citep{fouedjio2015estimation}, where $\ovec = (0, 0)'$. The data were simulated on a 101 $\times$ 101 grid defined on the geographic domain $G$. We randomly sampled without replacement 1000 locations from $G$ and used them as measurement locations.

\begin{figure}[t!]
	\centering
	\includegraphics[width=0.6\textwidth]{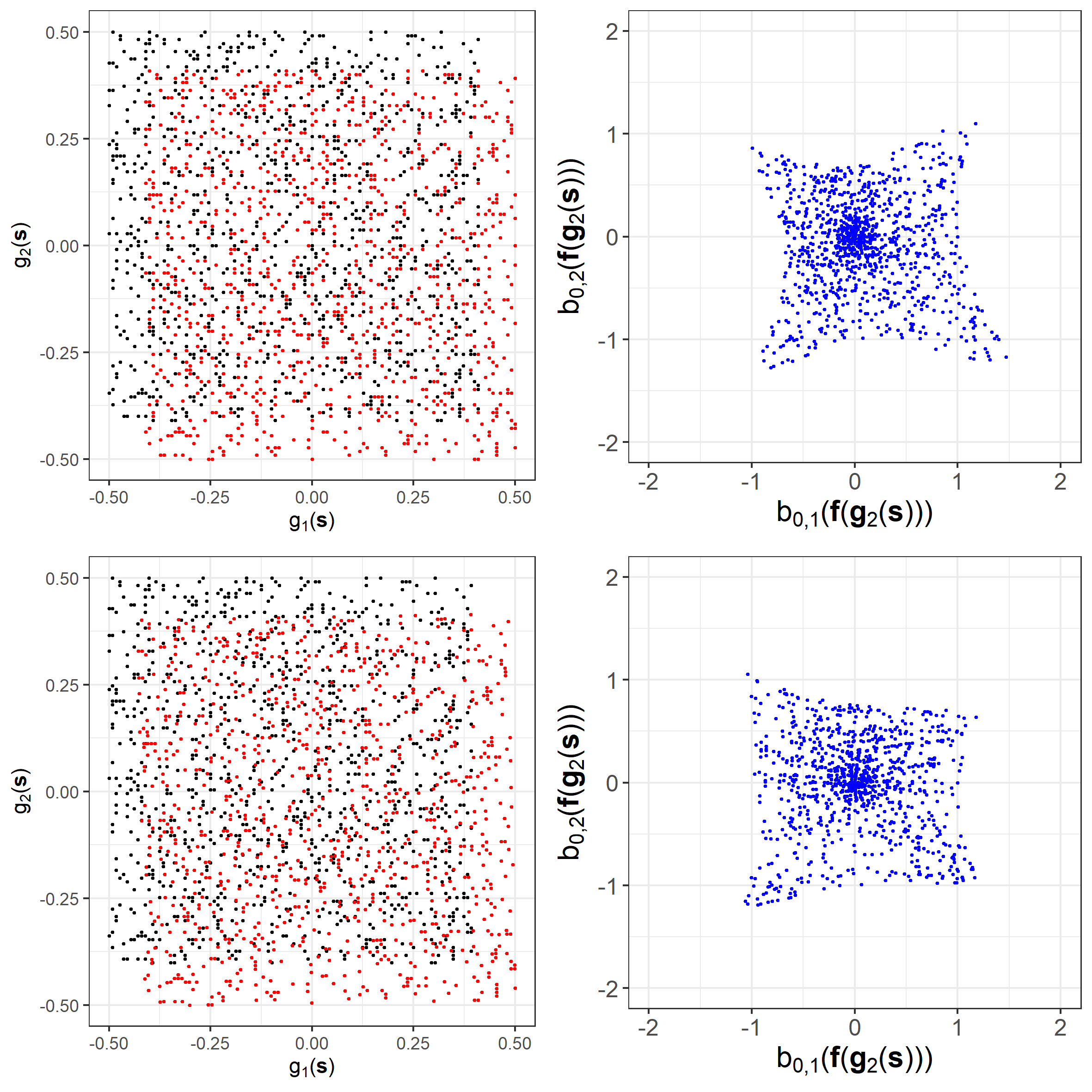}
	\caption{
		\footnotesize{Measurement locations under the aligning functions and the homogenized warping functions for the data set in Section \ref{sec:misspecified2}. Top row: Measurement locations under the true aligning function (left panel) and the true warping function (right panel). Bottom row: Measurement locations under the estimated aligning function (left panel) and the estimated warping function (right panel).}
	}
	\label{fig:warped_loc_misspecified2}
\end{figure}

As in Section \ref{sec:misspecified1}, we examined the ability of the DCSM to retrieve the true warping function using the same model as Model S4.2.5 in Section \ref{sec:simstudies_asym}. Figure \ref{fig:warped_loc_misspecified2} shows the measurement locations under the true aligning and warping function, and the measurement locations under the estimated aligning and warping function. We observe that the DCSM has correctly estimated the southeastern shift in the aligning function, and the contraction in the middle region of the warped domain.

\subsection{Simulated Trivariate Symmetric Nonstationary Data ~}\label{sec:trivar}

In the previous data illustrations, we considered bivariate spatial data. In this section, we consider trivariate spatial data. We simulated trivariate data from a Gaussian DCSM, $\tilde \Yvec(\cdot)$, with constant mean. The data were simulated on an equally spaced 51 $\times$ 51 grid of the geographic domain, $G \equiv [-0.5, 0.5] \times [-0.5, 0.5]$ resulting in 2601 data. The warping function we used was a composition of axial warping units, followed by a single-resolution radial basis function unit, followed by a M{\"o}bius transformation unit. On the warped domain, we modeled the covariances using a trivariate stationary, isotropic parsimonious Mat{\'e}rn model. As in Section \ref{sec:simstudies_sym}, we randomly sampled 1000 locations from the grid and used these as measurement locations.

\begin{figure}
	\centering
	\includegraphics[width=0.6\textwidth]{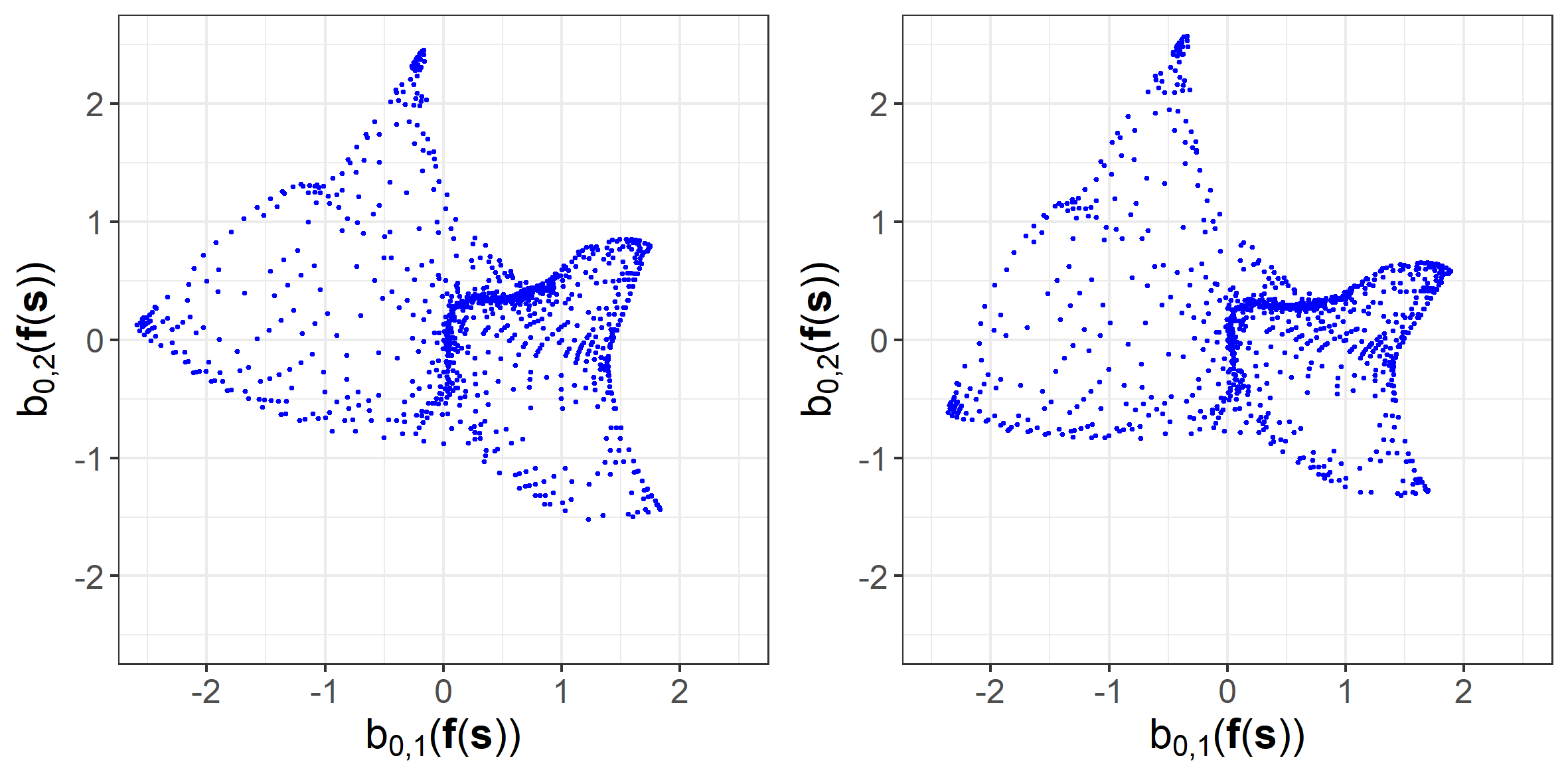}
	\caption{
		\footnotesize{Measurement locations under the true (left panel) and estimated (right panel) homogenized warping functions for the data set in Section \ref{sec:trivar}.}
	}
	\label{fig:warped_loc_trivar}
\end{figure}

We compared the predictive performance of the trivariate stationary parsimonious Mat{\'e}rn model (Model S4.4.1) to those of the trivariate DCSM (Model S4.4.2) by calculating the cross-validated predictive performance at the remaining 1601 locations using the RMSPE and the CRPS. Table \ref{tbl:trivar} summarizes the results from this study. From this table, it is clear that, similar to the bivariate case in Section \ref{sec:simstudies_sym}, there is a large improvement in RMSPE and CRPS when using the DCSM (Model S4.4.2) over the stationary parsimonious Mat{\'e}rn model (Model S4.4.1). The visualization of the estimated warping function is given in Figure \ref{fig:warped_loc_trivar}. We can see that, similar to the bivariate case, the estimated warping function has retrieved important features such as the contraction in the middle part of the domain. However, we find that more iterations are needed, thus more computing time, to train the DCSM in the trivariate-data example.

\begin{table}
	\centering
	\caption{Hold-out-validation results, AIC, and the time required to fit, for the simulation study in Section \ref{sec:trivar}, where the measurement locations are randomly sampled from $G = [-0.5, 0.5] \times [-0.5, 0.5]$.}
	\label{tbl:trivar}
	\bgroup
	\def\arraystretch{1}
	\begin{tabular}{ |c|c|c|c|c|c|c|c|c| }
		\hline
		& \multicolumn{2}{c|}{$\tilde Y_1(\cdot)$} & \multicolumn{2}{c|}{$\tilde Y_2(\cdot)$} & \multicolumn{2}{c|}{$\tilde Y_3(\cdot)$} & & \\
		\cline{2-7}
		& RMSPE & CRPS & RMSPE & CRPS & RMSPE & CRPS & AIC & Time (s) \\
		\hline
		Model S4.4.1 & 0.309 &	0.171  & 0.099 & 0.053 & 0.035 & 0.019 & -1924.0 & 1385.4 \\
		\hline
		Model S4.4.2 & 0.276 &	0.148  & 0.077 & 0.040 & 0.027 & 0.015 & -2865.0 & 9047.0  \\
		\hline
	\end{tabular}
	\egroup
\end{table}

\subsection{Modeling Maximum and Minimum Temperatures in Southwestern United States ~}\label{sec:maxmin}

In this section, we consider monthly maximum and minimum temperatures (which are calculated by averaging daily maximum and minimum temperatures over the month) in July 2018 at 909 weather stations over four southwestern states in the United States: Utah, Colorado, Arizona, and New Mexico. The data were extracted from the Global Summary of the Month data set of the National Oceanic and Atmospheric Administration\footnotetext{$^1$\url{https://data.nodc.noaa.gov/cgi-bin/iso?id=gov.noaa.ncdc:C00946}}\footnotemark{$^1$}.

We compared the performance of the bivariate DCSMs to that of bivariate stationary Mat{\'e}rn models. Elevation is a potential covariate when modeling temperature. We considered two trend models, namely one with only an intercept (in which case $Y_1(\cdot)$ and $Y_2(\cdot)$ are highly nonstationary) and one with an intercept and elevation as a covariate (in which case $Y_1(\cdot)$ and $Y_2(\cdot)$ can be expected to be nonstationary but less so). Also, the maximum and minimum temperatures in a given month (here July 2018) can be treated as symmetric spatial processes since their interaction can be expected to be highly co-located. Thus, asymmetry of the cross-covariance matrices was not modeled in this data illustration. We considered the following four models:
\begin{itemize}
	\item Model S4.5.1: A bivariate, stationary, parsimonious Mat{\'e}rn covariance model, and with only an intercept in the trend. 
	\item Model S4.5.2: A bivariate, symmetric DCSM with the parsimonious bivariate Mat{\'e}rn covariance on the warped domain, and with only an intercept in the trend
	\item Model S4.5.3: A bivariate, stationary, parsimonious Mat{\'e}rn covariance model, and with an intercept and elevation as trend. 
	\item Model S4.5.4: A bivariate, symmetric DCSM with the parsimonious bivariate Mat{\'e}rn covariance on the warped domain, and with an intercept and elevation as trend.
\end{itemize}
We used the same general architecture for the warping function of the bivariate DCSM as was used in the simulation study in Section \ref{sec:simstudies_sym}.

We first examined the predictive performance of the four models when the data were missing at random by performing a five-fold cross-validation. We randomly chose 905 stations from the 909 stations for the study, and randomly divided the 905 stations into five groups of 181 stations each in order to carry out a five-fold cross-validation.
Table \ref{tbl:crossvalid_temp} shows the results of this five-fold cross-validation study. We see a slight improvement in the predictive performance of the DCSMs over the corresponding stationary covariance models, on the order of 3--5\% for RMSPE and CRPS. We also see a substantial improvement when elevation is included in the trend model, showing that addressing nonstationarity in the mean function is helpful in this application.  

\begin{table}
	\centering
	\caption
	{
		Five-fold cross-validation results, AIC, and the time required to fit, for the maximum and minimum temperature data in the southwestern USA for the study in Section \ref{sec:maxmin}, where data were missing at random.
	}
	\label{tbl:crossvalid_temp}
	\bgroup
	\def\arraystretch{1}
	\begin{tabular}{ |c|c|c|c|c|c|c| }
		\hline
		& \multicolumn{2}{c|}{$T_{max}$} & \multicolumn{2}{c|}{$T_{min}$} & & \\
		\cline{2-5}
		& RMSPE & CRPS & RMSPE & CRPS & AIC & Time (s) \\
		\hline
		Model S4.5.1 & 3.17 & 1.74 & 2.44 & 1.33 & 7140.8 & 359.4 \\
		\hline
		Model S4.5.2 & 3.09 & 1.69 & 2.31 & 1.26 & 7003.5 & 791.8 \\
		\hline
		Model S4.5.3 & 1.07 & 0.58 & 2.02 & 1.14 & 5475.5 & 413.2 \\
		\hline
		Model S4.5.4 & 1.04 & 0.57 & 1.98 & 1.11 & 5398.5 & 766.3 \\
		\hline
	\end{tabular}
	\egroup
\end{table}

\begin{figure}
	\centering
	
	\includegraphics[width=0.8\textwidth]{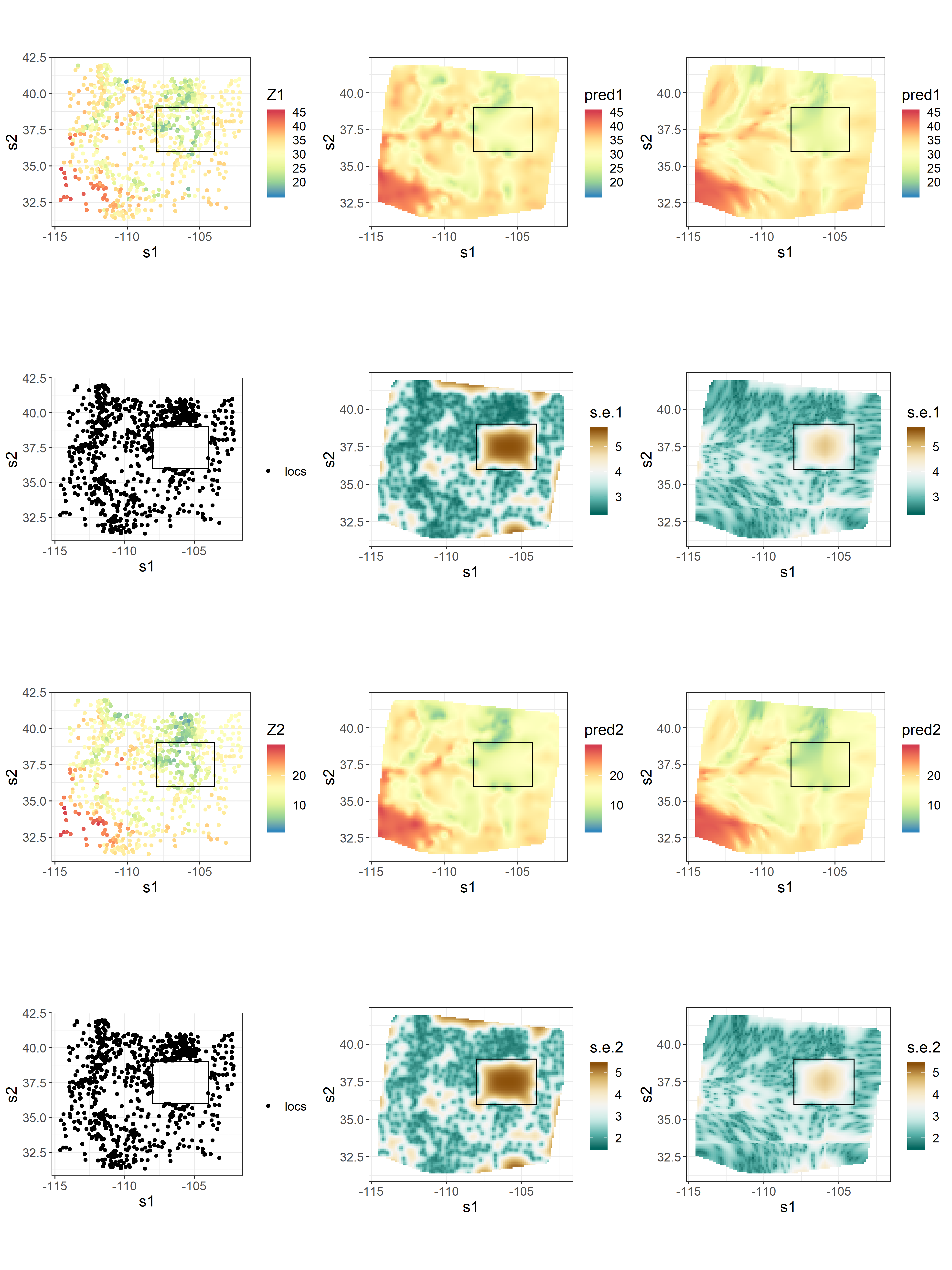}
	
	\caption{
		Comparison of predictions and prediction standard errors when using a bivariate symmetric, stationary, parsimonious Mat{\'e}rn model with constant mean (Model S4.5.1), and a bivariate symmetric DCSM with constant mean (Model S4.5.2). Data were left out of the region enclosed by the black rectangle.
		First row: Maximum temperature observations, $Z_1$ (left panel); predictions obtained using Model S4.5.1 (center panel), and Model S4.5.2 (right panel). Second row: Locations of the measurement of $Z_1$ (left panel); prediction standard errors obtained when using Model S4.5.1 (center panel), and Model S4.5.2 (right panel). Third and fourth rows: Analogous to the first and second rows, respectively, for the minimum temperature, $Z_2$.
	}
	\label{fig:weather_gap_study}
\end{figure}

We next considered hold-out validation, where the data are missing in a block, and we held out 131 stations lying between 36$^{\circ}$N--39$^{\circ}$N and 104$^{\circ}$W--108$^{\circ}$W from all the 909 stations. The western part of the hold-out region is a mountainous area that extends into the non-hold-out region, while the eastern part of the hold-out region is an area with lower elevation. We then fitted the four models to the remaining 778 stations. Table \ref{tbl:holdout_temp} displays the hold-out validation results for predicting the maximum and minimum temperatures at the 131 hold-out stations, and Figure \ref{fig:weather_gap_study} shows the maps of predictions and prediction standard errors for Model S4.5.1 and Model S4.5.2.

\begin{figure}
	\centering
	\includegraphics[width=0.6\textwidth]{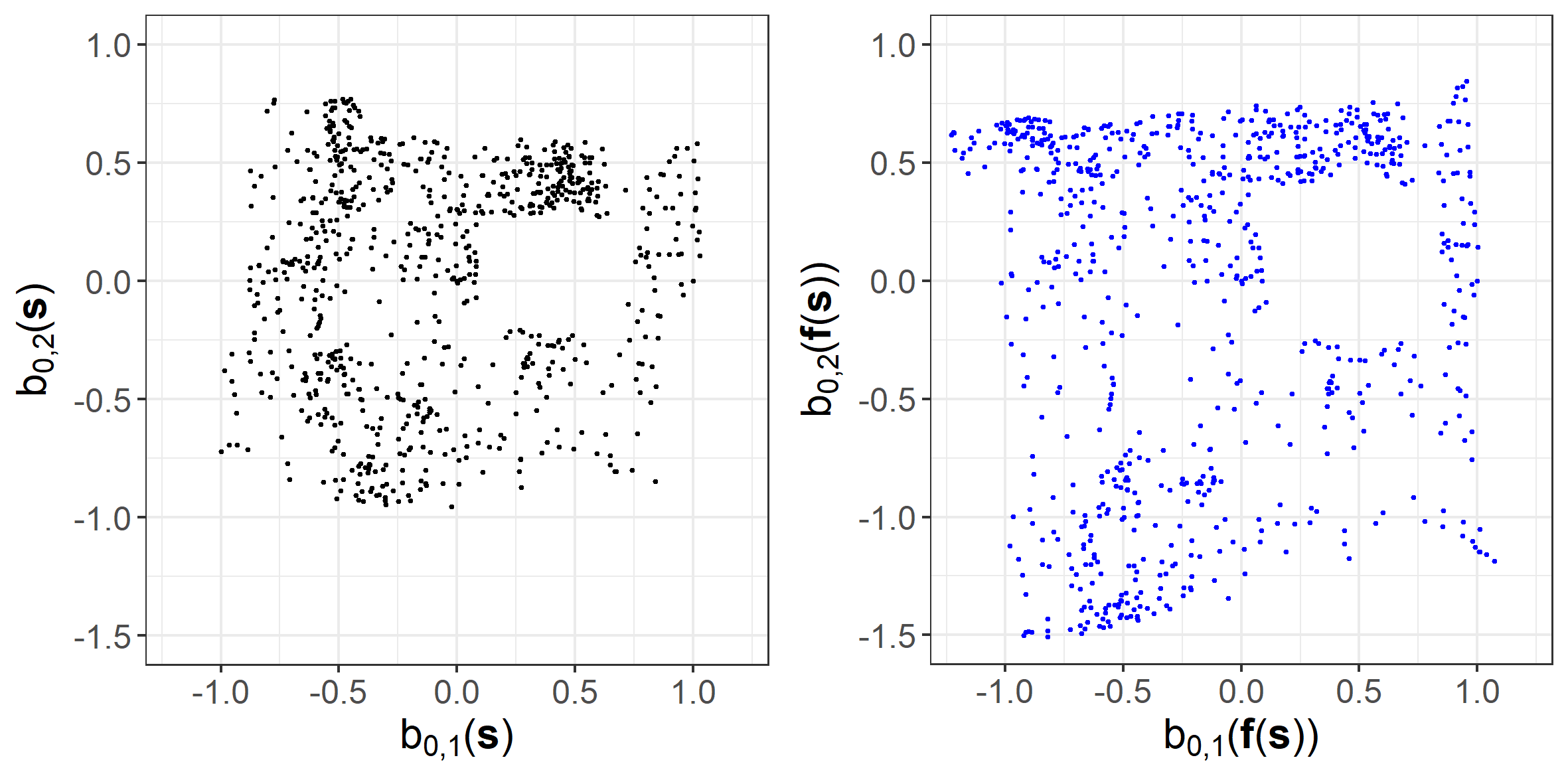}
	\caption{
		\footnotesize{Measurement locations on the geographical domain and under the estimated homogenized warping function for the data set in Section \ref{sec:maxmin}.}
	}
	\label{fig:maxmin_warping}
\end{figure}

From Table \ref{tbl:holdout_temp}, the predictive performances of the bivariate DCSMs show a huge improvement in RMSPE and CRPS, on the order of 30\% over those of the bivariate stationary covariance models when predicting maximum temperature. The visualization in Figure \ref{fig:weather_gap_study} further illustrates the utility of using DCSMs: With the intercept-only structure (i.e., constant mean) in the mean, the bivariate DCSM can predict lower temperatures on the western part of the hold-out region (that is, the mountainous areas), whereas the stationary model reverts to the constant mean. The DCSM also produces lower prediction standard errors over the hold-out region than the stationary model. These results corroborate those from the simulation study in Section \ref{sec:simstudies_sym}, which showed that bivariate DCSMs can be useful even when data are missing over a large region. Figure \ref{fig:maxmin_warping} shows the estimated warping function in Model S4.5.2.

\begin{table}
	\centering
	\caption{
		Hold-out-validation results, AIC, and the time required to fit, for the maximum and minimum temperature data in the southwestern USA for the study in Section \ref{sec:maxmin}, where data were missing in a block shown in Figure \ref{fig:weather_gap_study}.
	}
	\label{tbl:holdout_temp}
	\bgroup
	\def\arraystretch{1}
	\begin{tabular}{ |c|c|c|c|c|c|c| }
		\hline
		& \multicolumn{2}{c|}{$T_{max}$} & \multicolumn{2}{c|}{$T_{min}$} & & \\
		\cline{2-5}
		& RMSPE & CRPS & RMSPE & CRPS & AIC & Time (s) \\
		\hline
		Model S4.5.1 & 4.66 & 2.59 & 3.90 & 2.20 & 7610.1 & 418.7 \\
		\hline
		Model S4.5.2 & 3.84 &	2.19 & 2.57 & 1.51 & 7474.6 & 915.0 \\
		\hline
		Model S4.5.3 & 1.54 &	0.87 & 2.11 & 1.22 & 5858.4 & 475.0 \\
		\hline
		Model S4.5.4 & 1.06 &	0.59 & 1.99 & 1.16 & 5787.2 & 898.5 \\
		\hline
	\end{tabular}
	\egroup
\end{table}

\newpage

\bibliographystyle{apalike}
\bibliography{Bibliography}

\begin{thebibliography}{}

\bibitem[Allaire and Tang, 2019]{TensorflowR}
Allaire, J.~J. and Tang, Y. (2019).
\newblock {\em tensorflow: {R} Interface to `{T}ensor{F}low'}.
\newblock Online: Available from \url{https://github.com/rstudio/tensorflow}.

\bibitem[Anderes and Chatterjee, 2009]{anderes2009consistent}
Anderes, E. and Chatterjee, S. (2009).
\newblock Consistent estimates of deformed isotropic {G}aussian random fields
  on the plane.
\newblock {\em The Annals of Statistics}, 37:2324--2350.

\bibitem[Anderes and Stein, 2008]{anderes2008estimating}
Anderes, E.~B. and Stein, M.~L. (2008).
\newblock Estimating deformations of isotropic {G}aussian random fields on the
  plane.
\newblock {\em The Annals of Statistics}, 36:719--741.

\bibitem[Apanasovich and Genton, 2010]{apanasovich2010cross}
Apanasovich, T.~V. and Genton, M.~G. (2010).
\newblock Cross-covariance functions for multivariate random fields based on
  latent dimensions.
\newblock {\em Biometrika}, 97:15--30.

\bibitem[Apanasovich et~al., 2012]{apanasovich2012valid}
Apanasovich, T.~V., Genton, M.~G., and Sun, Y. (2012).
\newblock A valid {M}at{\'e}rn class of cross-covariance functions for
  multivariate random fields with any number of components.
\newblock {\em Journal of the American Statistical Association}, 107:180--193.

\bibitem[Calandra et~al., 2016]{Calandra_2016}
Calandra, R., Peters, J., Rasmussen, C.~E., and Deisenroth, M.~P. (2016).
\newblock Manifold {G}aussian processes for regression.
\newblock In {\em Proceedings of the 2016 International Joint Conference on
  Neural Networks (IJCNN)}, pages 3338--3345. IEEE, Vancouver, BC, Canada.

\bibitem[Cressie and Johannesson, 2008]{cressie2008fixed}
Cressie, N. and Johannesson, G. (2008).
\newblock Fixed rank kriging for very large spatial data sets.
\newblock {\em Journal of the Royal Statistical Society: Series B},
  70:209--226.

\bibitem[Cressie and Lahiri, 1993]{cressie1993asymptotic}
Cressie, N. and Lahiri, S.~N. (1993).
\newblock The asymptotic distribution of {REML} estimators.
\newblock {\em Journal of Multivariate Analysis}, 45:217--233.

\bibitem[Cressie and Lahiri, 1996]{cressie1996asymptotics}
Cressie, N. and Lahiri, S.~N. (1996).
\newblock Asymptotics for {REML} estimation of spatial covariance parameters.
\newblock {\em Journal of Statistical Planning and Inference}, 50:327--341.

\bibitem[Cressie and Zammit-Mangion, 2016]{cressie2016multivariate}
Cressie, N. and Zammit-Mangion, A. (2016).
\newblock Multivariate spatial covariance models: {A} conditional approach.
\newblock {\em Biometrika}, 103:915--935.

\bibitem[Damian et~al., 2001]{damian2001bayesian}
Damian, D., Sampson, P.~D., and Guttorp, P. (2001).
\newblock Bayesian estimation of semi-parametric non-stationary spatial
  covariance structures.
\newblock {\em Environmetrics}, 12:161--178.

\bibitem[Fouedjio et~al., 2015]{fouedjio2015estimation}
Fouedjio, F., Desassis, N., and Romary, T. (2015).
\newblock Estimation of space deformation model for non-stationary random
  functions.
\newblock {\em Spatial Statistics}, 13:45--61.

\bibitem[Fuglstad et~al., 2015]{fuglstad2015exploring}
Fuglstad, G.-A., Lindgren, F., Simpson, D., and Rue, H. (2015).
\newblock Exploring a new class of non-stationary spatial {G}aussian random
  fields with varying local anisotropy.
\newblock {\em Statistica Sinica}, 25:115--133.

\bibitem[Gelfand et~al., 2004]{gelfand2004nonstationary}
Gelfand, A.~E., Schmidt, A.~M., Banerjee, S., and Sirmans, C. (2004).
\newblock Nonstationary multivariate process modeling through spatially varying
  coregionalization.
\newblock {\em Test}, 13:263--312.

\bibitem[Gneiting et~al., 2010]{gneiting2010matern}
Gneiting, T., Kleiber, W., and Schlather, M. (2010).
\newblock Mat{\'e}rn cross-covariance functions for multivariate random fields.
\newblock {\em Journal of the American Statistical Association},
  105:1167--1177.

\bibitem[Gneiting and Raftery, 2007]{gneiting2007strictly}
Gneiting, T. and Raftery, A.~E. (2007).
\newblock Strictly proper scoring rules, prediction, and estimation.
\newblock {\em Journal of the American Statistical Association}, 102:359--378.

\bibitem[Goulard and Voltz, 1992]{goulard1992linear}
Goulard, M. and Voltz, M. (1992).
\newblock Linear coregionalization model: {T}ools for estimation and choice of
  cross-variogram matrix.
\newblock {\em Mathematical Geology}, 24:269--286.

\bibitem[Heaton et~al., 2019]{heaton2019case}
Heaton, M.~J., Datta, A., Finley, A.~O., Furrer, R., Guinness, J., Guhaniyogi,
  R., Gerber, F., Gramacy, R.~B., Hammerling, D., Katzfuss, M., Lindgren, F.,
  Nychka, D.~W., Sun, F., and Zammit-Mangion, A. (2019).
\newblock A case study competition among methods for analyzing large spatial
  data.
\newblock {\em Journal of Agricultural, Biological and Environmental
  Statistics}, 24:398--425.

\bibitem[Higdon et~al., 1999]{higdon1999non}
Higdon, D., Swall, J., and Kern, J. (1999).
\newblock Non-stationary spatial modeling.
\newblock {\em Bayesian Statistics}, 6:761--768.

\bibitem[Hildeman et~al., 2019]{hildeman2019joint}
Hildeman, A., Bolin, D., and Rychlik, I. (2019).
\newblock Joint spatial modeling of significant wave height and wave period
  using the {SPDE} approach.
\newblock {\em arXiv preprint, arXiv:1906.00286}.

\bibitem[Hu and Steinsland, 2016]{Hu_2016}
Hu, X. and Steinsland, I. (2016).
\newblock Spatial modeling with system of stochastic partial differential
  equations.
\newblock {\em Wiley Interdisciplinary Reviews: Computational Statistics},
  8:112--125.

\bibitem[Kadane, 1974]{kadane1974role}
Kadane, J.~B. (1974).
\newblock The role of identification in {B}ayesian theory.
\newblock In Fienberg, S.~E. and Zellner, A., editors, {\em Studies in Bayesian
  Econometrics and Statistics}, pages 175--191. Amsterdam, The Netherlands.

\bibitem[Kleiber and Nychka, 2012]{kleiber2012nonstationary}
Kleiber, W. and Nychka, D. (2012).
\newblock Nonstationary modeling for multivariate spatial processes.
\newblock {\em Journal of Multivariate Analysis}, 112:76--91.

\bibitem[Li and Zhang, 2011]{li2011approach}
Li, B. and Zhang, H. (2011).
\newblock An approach to modeling asymmetric multivariate spatial covariance
  structures.
\newblock {\em Journal of Multivariate Analysis}, 102:1445--1453.

\bibitem[Lindgren et~al., 2011]{lindgren2011explicit}
Lindgren, F., Rue, H., and Lindstr{\"o}m, J. (2011).
\newblock An explicit link between {G}aussian fields and {G}aussian {M}arkov
  random fields: the stochastic partial differential equation approach.
\newblock {\em Journal of the Royal Statistical Society: Series B},
  73:423--498.

\bibitem[Majumdar and Gelfand, 2007]{majumdar2007multivariate}
Majumdar, A. and Gelfand, A.~E. (2007).
\newblock Multivariate spatial modeling for geostatistical data using convolved
  covariance functions.
\newblock {\em Mathematical Geology}, 39:225--245.

\bibitem[Meiring et~al., 1997]{Meiring_1997}
Meiring, W., Monestiez, P., Sampson, P., and Guttorp, P. (1997).
\newblock Developments in the modelling of nonstationary spatial covariance
  structure from space-time monitoring data.
\newblock In Baafi, E.~Y. and Schofield, N.~A., editors, {\em Geostatistics
  Wollongong `96}, pages 162--173. Kluwer, Dordrecht, The Netherlands.

\bibitem[Messick et~al., 2017]{messick2017multivariate}
Messick, R.~M., Heaton, M.~J., and Hansen, N. (2017).
\newblock Multivariate spatial mapping of soil water holding capacity with
  spatially varying cross-correlations.
\newblock {\em Annals of Applied Statistics}, 11:69--92.

\bibitem[Nguyen et~al., 2017]{Nguyen_2017}
Nguyen, H., Cressie, N., and Braverman, A. (2017).
\newblock Multivariate spatial data fusion for very large remote sensing
  datasets.
\newblock {\em Remote Sensing}, 9:142.

\bibitem[Olea and Pardo-Iguzquiza, 2011]{olea2011generalized}
Olea, R.~A. and Pardo-Iguzquiza, E. (2011).
\newblock Generalized bootstrap method for assessment of uncertainty in
  semivariogram inference.
\newblock {\em Mathematical Geosciences}, 43:203--228.

\bibitem[Paciorek and Schervish, 2006]{paciorek2006spatial}
Paciorek, C.~J. and Schervish, M.~J. (2006).
\newblock Spatial modelling using a new class of nonstationary covariance
  functions.
\newblock {\em Environmetrics}, 17:483--506.

\bibitem[Perrin and Monestiez, 1999]{perrin1999modelling}
Perrin, O. and Monestiez, P. (1999).
\newblock Modelling of non-stationary spatial structure using parametric radial
  basis deformations.
\newblock In Gomez-Hernandez, J., Soares, A., and Froidevaux, R., editors, {\em
  geoENV II—Geostatistics for Environmental Applications}, pages 175--186.
  Springer, New York, NY.

\bibitem[Qadir and Sun, 2020]{qadir2019semiparametric}
Qadir, G.~A. and Sun, Y. (2020).
\newblock Semiparametric estimation of cross-covariance functions for
  multivariate random fields.
\newblock {\em Biometrics}, in press,
  DOI:\url{https://doi.org/10.1111/biom.13323}.

\bibitem[Sampson et~al., 2001]{Sampson_2001}
Sampson, P., Damian, D., and Guttorp, P. (2001).
\newblock Advances in modeling and inference for environmental processes with
  nonstationary spatial covariance.
\newblock In Monestiez, P., Allard, D., and Froidevaux, R., editors, {\em
  GeoENV III--Geostatistics for Environmental Applications}, pages 17--32.
  Springer, New York, NY.

\bibitem[Sampson and Guttorp, 1992]{sampson1992nonparametric}
Sampson, P.~D. and Guttorp, P. (1992).
\newblock Nonparametric estimation of nonstationary spatial covariance
  structure.
\newblock {\em Journal of the American Statistical Association}, 87:108--119.

\bibitem[Schmidt and O'Hagan, 2003]{schmidt2003bayesian2}
Schmidt, A.~M. and O'Hagan, A. (2003).
\newblock Bayesian inference for non-stationary spatial covariance structure
  via spatial deformations.
\newblock {\em Journal of the Royal Statistical Society: Series B},
  65:743--758.

\bibitem[Solow, 1985]{solow1985bootstrapping}
Solow, A.~R. (1985).
\newblock Bootstrapping correlated data.
\newblock {\em Mathematical Geology}, 17:769--775.

\bibitem[Ver~Hoef and Barry, 1998]{ver1998constructing}
Ver~Hoef, J.~M. and Barry, R.~P. (1998).
\newblock Constructing and fitting models for cokriging and multivariable
  spatial prediction.
\newblock {\em Journal of Statistical Planning and Inference}, 69:275--294.

\bibitem[Ver~Hoef and Cressie, 1993]{ver1993multivariable}
Ver~Hoef, J.~M. and Cressie, N. (1993).
\newblock Multivariable spatial prediction.
\newblock {\em Mathematical Geology}, 25:219--240.

\bibitem[Wackernagel, 2003]{wackernagel2003multivariate}
Wackernagel, H. (2003).
\newblock {\em Multivariate Geostatistics: An Introduction with Applications}.
\newblock Springer, Berlin.

\bibitem[Wiens et~al., 2020]{wiens2019surface}
Wiens, A., Kleiber, W., Barnhart, K.~R., and Sain, D. (2020).
\newblock Surface estimation for multiple misaligned point sets.
\newblock {\em Mathematical Geosciences}, 52:527--542.

\bibitem[Zammit-Mangion et~al., 2015a]{zammit2015spatio}
Zammit-Mangion, A., Cressie, N., Ganesan, A.~L., O'Doherty, S., and Manning,
  A.~J. (2015a).
\newblock Spatio-temporal bivariate statistical models for atmospheric
  trace-gas inversion.
\newblock {\em Chemometrics and Intelligent Laboratory Systems}, 149:227--241.

\bibitem[Zammit-Mangion et~al., 2019]{azm2019deep}
Zammit-Mangion, A., Ng, T. L.~J., Vu, Q., and Filippone, M. (2019).
\newblock Deep compositional spatial models.
\newblock {\em arXiv preprint, arXiv:1906.02840}.

\bibitem[Zammit-Mangion et~al., 2015b]{zammit2015multivariate}
Zammit-Mangion, A., Rougier, J., Sch{\"o}n, N., Lindgren, F., and Bamber, J.
  (2015b).
\newblock Multivariate spatio-temporal modelling for assessing {A}ntarctica's
  present-day contribution to sea-level rise.
\newblock {\em Environmetrics}, 26:159--177.

\bibitem[Zhang, 2004]{zhang2004inconsistent}
Zhang, H. (2004).
\newblock Inconsistent estimation and asymptotically equal interpolations in
  model-based geostatistics.
\newblock {\em Journal of the American Statistical Association}, 99:250--261.

\end{thebibliography}


\end{document}